# Are particles self-organized systems?

**Vladimir A. Manasson**
Correspondence: v.a.manasson@cox.net

**Abstract:**

Where did elementary particles come from? What mechanisms are responsible for their occurrence and maintenance? Are they compound or truly elementary? Is vacuum primordial soup where elementary particles are born? Are quantum behavior and relativistic phenomena fundamental or emergent? This paper describes a primitive active medium far from thermodynamic equilibrium, which we associate with vacuum and in which a system of particles and fields arises, similar to that described by the standard model. Phenomena usually attributed to quantum or relativistic media emerge during vacuum self-organization. These include discrete spectra of ground states, charges, oscillation periods, and link flavors, spatial phase coherency, virtual states, tunneling, entanglement, time-related uncertainty of states, and coexistent Planck-like and Einstein-like time scales. The form of vacuum self-organization is a coherent time-crystal network. Here, different fields arise as quantum states of the same gravitation-antigravitation field. Particle-like entities emerge as topological defects. The analysis was accompanied by numerical estimates of several physical constants, the values of which had never previously been derived from theory. These results are consistent with the experimental data.

**Keywords:** open dynamical systems, self-organization, nonlinear forces, asymptotic stability, temporal periodicity, time crystals, attractor topology, period-doubling phase transitions, synchronization, quantization, time dilation, particles and fields, Feigenbaum universality, physical constants, time metrics.

## 1. Introduction

The particles of the standard model (SM) are accepted as elementary building blocks for all matter in the universe. However, their number, diverse properties and complex interactions call into question their elementary nature. SM does not indicate their origin and how they maintain and reproduce their properties. The current situation with elementary particles is reminiscent of the pre-Darwinian situation with biological species. The species were systematized, and their interrelations were generally understood, but their origin and diversity have remained unclear. References to random occurrence in multiverse or other environments, anthropic principles, or intelligent design leave the questions open. Turning to more complex systems and environments than the SM takes us further away from the answer.

The essence of Darwinian evolution is the self-organization of an environment in which random events are natural. What if we apply this approach to elementary particles? In the following, we have shown how, in a primitive random environment, which we identify with a physical vacuum, a system arises that resembles the SM in its complexity, organization, and even some quantitative parameters.

We mainly considered the temporal aspects of medium self-organization. We turned to its simplest form: temporal periodicity. The latter relates the medium to the time crystals (TCs) (the term was coined by Winfrey [1] and revived by Wilczek [2,3]). Inherently dynamic, they cannot exist in a frozen environment [4-6]. The TCs controlled by external periodic signals [7-10] are not self-sustaining; therefore, they are also beyond our interest. The TCs we will discuss here are known as self-oscillators, generators, orbitally stable limit cycles, vortices, rotators, etc. For brevity, we call them oscillators, keeping in mind that they are dissipative, self-organized and self-sustained. Nature and technology give us numerous examples of such systems: von Kármán vortices, color-changing chemical reactions, living cells, biological organs (hearts, lungs, intestines, brains), songbirds, cicada choruses, organ pipes, electronic signal generators, lasers, etc.

Born in a noisy environment, a self-sustained oscillator constantly experiences disturbances that transfer it from one evolutionary trajectory to another. However, thanks to dissipation, the system tends to return to the initial state of dynamic equilibrium. We call this state the attractor or fixed point, and we call this tendency asymptotic stability. It is interesting that elementary particles possess this property. We can take an electron as an example. In physics, the electron charge is one of the most important fundamental constants. However, in reality, the charge is not constant. When the particle is perturbed, its charge can change substantially. The only true constant is the value that the electron possesses in equilibrium. This value is a fixed point at which the electron returns regardless of its previous history, the details of its origin, or the level of excitation. The vacuum itself represents an asymptotically stable dissipative medium. This is manifested in the fact that all excitations (virtual particles) die within the time allotted to them by the energy-time uncertainty principle.

In addition to periodicity, we will also consider other TC properties reminiscent of those we can find in regular crystals. We will encounter temporal symmetries, temporal dislocations, collective excitation in the form of waves, spinorial states, defect transfers, etc. Perhaps most important are the *temporal* phase transitions, which essentially change the temporal symmetries and properties of TCs.

In sections 2–5, we discuss the emergent properties of TCs, including those that are usually assigned a priori to quantum or relativistic media. In the subsequent sections, we provide a model of vacuum self-organization and compare the patterns that emerge in this medium with the fields and particles of the SM.

Numerical simulations were carried out using the formalism of iterated maps [11–14], which is well established as an effective instrument for the exploration of self-organized media.

Some ideas and results relevant to this study have been published by the author in [15] and a few preprints. However, the material presented in this article is self-sufficient.

## 2. Parameter $\mathcal{E}$ as a measure of asymptotic stability

A typical evolutionary trajectory of a self-sustained oscillator in the state space is a curve asymptotically approaching an exponential spiral, which converges toward an attractor (a closed curve). The spirals can be parameterized by two simultaneous motions described by imaginary and real exponents:

$$\begin{cases} \varphi(t) = \varphi_0 \, exp(i2\pi t/\mathcal{T}) \\ r(t) = r_0 \, exp(-t/\tau) \end{cases}, \tag{1}$$

where $\varphi$ is the rotation angle, $r$ is the distance between the spiral and its attractor, $\mathcal{T}$ is the rotation period, and $\tau$ is the relaxation (dissipation) time constant.

The evolution trajectories converging to the same attractor form the attractor basin. Remarkably, they all converge with approximately the same value of $\tau$. Therefore, the parameter $\tau$ can be used as a measure for comparing attractors. The attractors whose spirals have a minimal $\tau$ are called superattractors. These are the oscillator ground states.

In the following analysis, as a measure of asymptotic stability, we use the parameter $\mathcal{E}$, which we define as

$$\mathcal{E} = -\tau^{-1}, \tag{2}$$

where $\tau$ is the respective dissipation time constant. As we will see below, $\mathcal{E}$ bears some resemblance to energy. It is expressed in units of frequency and it is minimal in the ground states. It is linked to the dimensionless Lyapunov exponent $\lambda$ as the "action integral" of $\mathcal{E}$ accumulated over one oscillation period $\mathcal{T}$:

$$\lambda = \int_t^{t+\mathcal{T}} \mathcal{E} \, dt = \mathcal{E}\mathcal{T}. \tag{3}$$

The second equality in (3) assumes that during one cycle, $\mathcal{E}$ is a constant. In the ground states, the “action” $\lambda$ is also minimal.

In a turbulent environment, perturbations randomly move oscillators from one trajectory to another. The frequency of these jumps increases with increasing environmental temperature. Below, we use a temperature scale that defines temperature $\theta$ as the frequency of these jumps:

$$\theta = t_\theta^{-1}. \tag{4}$$

where $t_\theta$ is the average time during which the oscillator is continuously on the same trajectory. $\theta$ is expressed in the same units as $\mathcal{E}$.

Let $v_\mathcal{E}(t)$ be the volume of the state space occupied by all attractor $\mathcal{E}$ trajectories at time $t$ after the beginning of evolution. The initial volumes $v_\mathcal{E}(t = 0)$ of different attractors overlap, which results in uncertainty $\Delta\mathcal{E}$ of parameter $\mathcal{E}$. $v_{\mathcal{E}}(t = 0)$ can be as large as the entire phase space volume $v_0$, then $\Delta\mathcal{E} \rightarrow \infty$. However, the volumes $v_{\mathcal{E}}$ decrease with time (with the same time constants as their spirals), and $\Delta\mathcal{E}$ also decreases without limitation until the final states are fully resolved, $\Delta\mathcal{E} \rightarrow 0$ at $t \rightarrow \infty$. This resembles the behavior of a quantum system subject to the uncertainty principle.

At any finite temperature $\theta$, the observation time is limited by the interval $t_\theta$, and accordingly, the volume $v_{\mathcal{E}}$ is compressed to the value

$$v_{\mathcal{E}}(t_\theta) = \frac{v_0}{exp(t_\theta \mathcal{E})}. \quad (5)$$

Assuming that the number of states (trajectories) in each volume is fixed and that the probability $w_{\mathcal{E}}$ of finding an oscillator in state $\mathcal{E}$ is proportional to the density of states, the probability

$$w_{\mathcal{E}} \propto v_{\mathcal{E}}(t_\theta)^{-1} = exp(-\mathcal{E}/\theta). \quad (6)$$

Expression (6) can be used to estimate the distribution of oscillators over $\mathcal{E}$. It has the form of the Boltzmann factor. Below, we will see how synchronization between oscillators turns this "classical" distribution into "quantum" distributions.

## 3. Connected time crystals

The word "phase" is used in two different senses: as a state of existence (a topological phase, vacuum phase, or phase transition) and as a fraction of a cycle in a periodic process (a phase difference $\Delta\varphi$, phase shift, or phase entrainment). Usually, the meaning is clear from the context; otherwise, it will be provided.

Weakly coupled oscillators evolve asynchronously. They have different periods and disconnected phases. A phase shift imposed on one of them does not change the dynamics of the others. This independence is lost when the oscillators interact more strongly and synchronize their dynamics. Synchronization is a phase transition that can take various forms [16-18]. Below, we will consider two distinct examples: phase entrainment and amplitude quenching.

Phase entrainment (also known as phase locking) is a ubiquitous phenomenon that can occur between oscillators of the same type, as well as between oscillators that produce different waveforms (regular or chaotic), have different designs and even use different principles of operation. Phase locking transforms coupled oscillators into a coherent network, where the phases of the oscillators are tied to each other and (after the synchronization process is stabilized) the phase difference $\psi$ between oscillators emerges as a new order parameter. Phase differences $\psi$ over the network form a field, which we call the $\psi$-field (or phase difference field).

Phase-locked oscillators also lose frequency independence. After synchronization, their arbitrary values from the continuous spectrum become interdependent and form a discrete spectrum or (in most cases) a single value (which we take as default). Any of the emergent periods can be used for setting a time scale common to the entire network. This global time scale is inherently quantum and can be compared with the Plank time.

We model link evolutionary dynamics using the Adler equation [19], which we have adapted [15] to the formalism of iterated maps:

$$\Delta\varphi_{n+1} = \Delta\varphi_n + \mathcal{T}\Delta\omega - \kappa sin\Delta\varphi_n. \qquad (7)$$

Here, $\mathcal{T}$ is the oscillator cycle duration after synchronization; $\Delta\varphi_{n+1}$ and $\Delta\varphi_n$ are the phase differences between the oscillators at times $t_{n+1}$ and $t_n$, separated by one cycle (time interval $\mathcal{T}$); $\Delta\omega$ is the difference between the oscillator frequencies before synchronization (natural frequencies); and $\kappa$ is the engagement coefficient, which characterizes the coupling strength between the oscillators and varies from $\kappa = 0$ (uncoupled) to $\kappa = 1$ (fully engaged).

Examples of phase difference evolution trajectories calculated using iterated map (7) are shown in Figure 1(a). They illustrate how an initially extended interval of states narrows down to a pointlike state. Regardless of the initial phase differences $\Delta\varphi_0$, which are spread across a $2\pi$ interval, the trajectories evolve to a fixed point (or attractor) $\psi = \Delta\varphi_\infty$. The position of the latter depends on the system prehistory (parameters $\Delta\omega$ and $\kappa$). After some initial evolution, the trajectories asymptotically become exponential (Figure 1(b)). Therefore, the parameters $\tau$, $\mathcal{E}$, $\lambda$ and $\theta$ introduced in the previous section also apply to oscillator links.

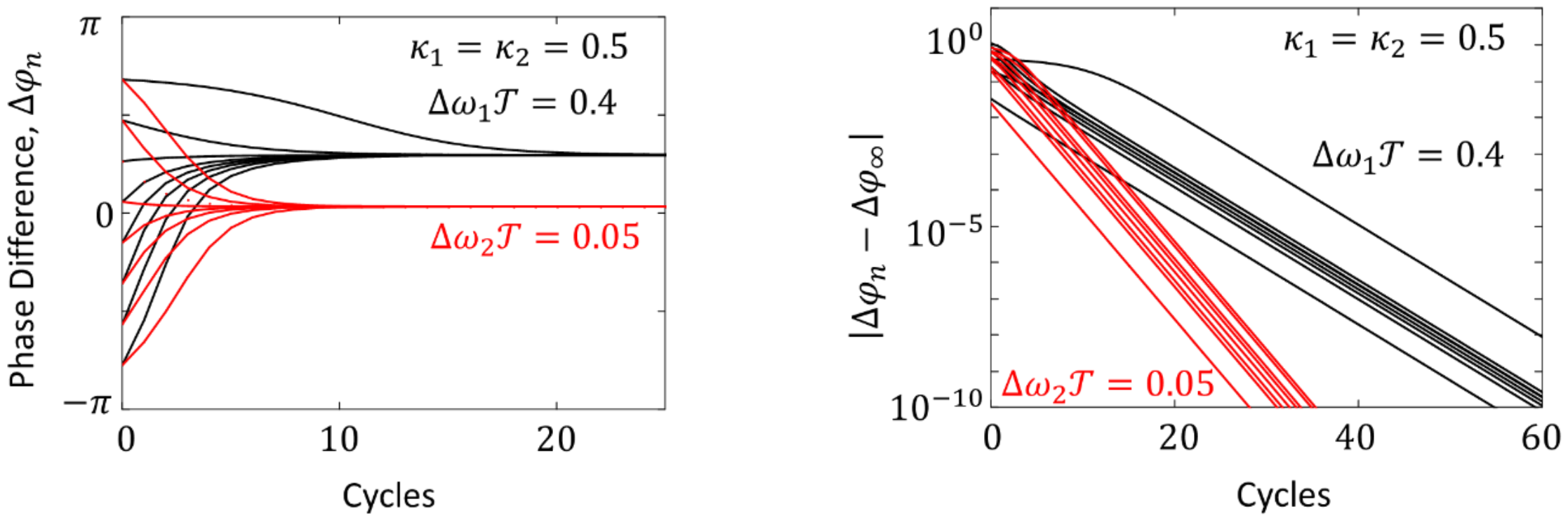

**(a)** **(b)**

**Figure 1.** Phase difference evolution trajectories: (**a**) phase differences $\Delta\varphi_n$ after the $n$th cycle; (**b**) evolutions of the differences between the phases $\Delta\varphi_n$ and the fixed points $\psi = \Delta\varphi_\infty$.

Examples of $\mathcal{E}(\psi, \kappa)$ plots simulated using iterated map (7) are shown in Figure 2. All ground states (minima) correspond to the phase difference $\psi = 0$ established after synchronization.

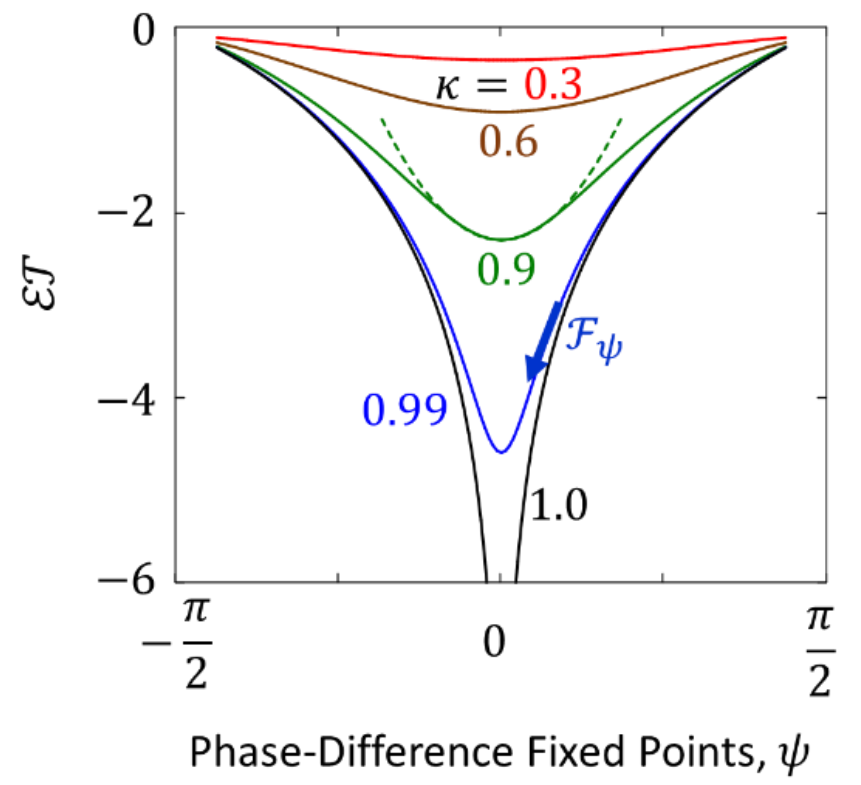

**Figure 2.** $\mathcal{E}(\psi)$ for links with different $\kappa$ values (solid lines). An approximation by a quadratic parabola (dashed line).

Like oscillators, oscillator links tend to stay statistically close to their ground state(s). The ambient noise pushes them out of this state. However, they keep returning to the ground state due to the higher probability of being there. This can be attributed to the action of an entropic restoring force $\mathcal{F}_\psi$ acting on the oscillator links. We call this force elastic. We define it as

$$\mathcal{F}_\psi = -d\mathcal{E}/d\psi, \quad (8)$$

Elastic forces form an $\mathcal{F}_\psi$-field in the network. The ground state is one where $\mathcal{F}_\psi = 0$. It happens at the point where phase difference $\psi = 0$.

Because of the convergence process, the network reaction to perturbations is not instantaneous. The delay plays the role of inertia. As a result, the elastic forces form phase-difference waves similar to those described in [20]. We call them $\psi$-waves.

Near the ground states, the dependences $\mathcal{E}(\psi)$ can be approximated by quadratic parabolas (an example is shown by the green dotted line at the bottom of the $\kappa = 0.9$ curve in Figure 2):

$$\mathcal{E} = (s/2)\psi^2 - \mathcal{E}_0, \quad (9)$$

where $\mathcal{E}_0$ is the ground state value and $s = \kappa/(1-\kappa)$ is a parameter that we call the stiffness. For weak engagement, when $\kappa \ll 1$, $s \approx \kappa$. For stronger engagements, $s$ increases faster than $\kappa$, and for full engagement, when $\kappa = 1$, $s \to \infty$.

Near the ground state, the elastic force is proportional to the phase difference $\psi$:

$$\mathcal{F}_\psi = -\, s\psi, \quad (10)$$

and $\psi$-waves are harmonic. Equation (10) plays the role of Hooke's law.

The nonlinearity arises with the link departure from the ground state. This is illustrated by the curves shown in Figure 3 in normalized coordinates. Hooke's growth is replaced by a sublinear region, then saturation, and, finally (at sufficiently large $\kappa$), a falling branch. As $\kappa$ increases, the positions of the curve maxima approach the ground state, $\mathcal{E}_0$, and at $\kappa = 1$, the ascending segment and saturation disappear, and only the descending branch remains. Force softening, described by descending branches, resembles the behavior of color forces in chromodynamics, a phenomenon known as asymptotic freedom.

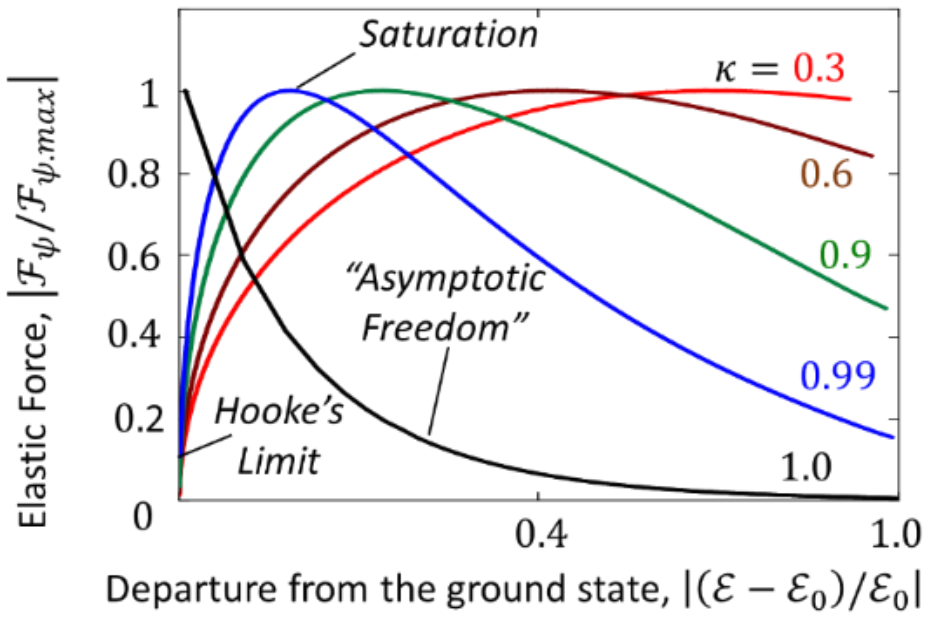

**Figure 3.** Normalized elastic force vs. normalized parameter $\mathcal{E}$.

Another nonlinear phenomenon that can occur in oscillator networks is solitary phase waves [21].

Dissipative systems are time irreversible. Irreversible time cannot be smoothly transformed into reversible space and vice versa. Lack of Lorentz invariance does not prevent time-crystal media from experiencing time dilation. This occurs due to local disturbance of the network and a corresponding increase in the dissipation time constant $\tau$ and leads to a slowdown of all local processes occurring in the network. The time dilation can be different at different locations. It can increase from $\tau_0 = \tau_{min}$ at the ground states to infinity at the states of unstable equilibrium ($\mathcal{E} = 0$). It can be used to set a local time scale, which can be associated with proper Einstein time in general relativity.

In Figure 4, we compare the time-crystal time dilation $\tau/\tau_0$ caused by network excitation with that of its relativistic analog $t/t_0$ caused by the presence of a black hole. The dependences of these two parameters on excitation are shown by the red and blue curves, respectively. Zero subindexes denote the ground states, which in the case of TCs corresponds to $\psi = 0$, and marks the flat space-time in the relativistic case. As a relativistic system, we choose a clock in the gravitational field of a black hole. The black hole Schwarzschild radius is $r_S$. As the excitation parameter, we use the ratio $r_S/R$, where $R$ is the distance between the clock and the black hole horizon. The ground state occurs when $R = \infty$ or $r_S/R = 0$. The time dilation $t/t_0$ as a function of $r_S/R$ was extracted as a time-related term from the Schwarzschild metric and is given by

$$t/t_0 = (1 - r_S/R)^{-1/2}. \tag{11}$$

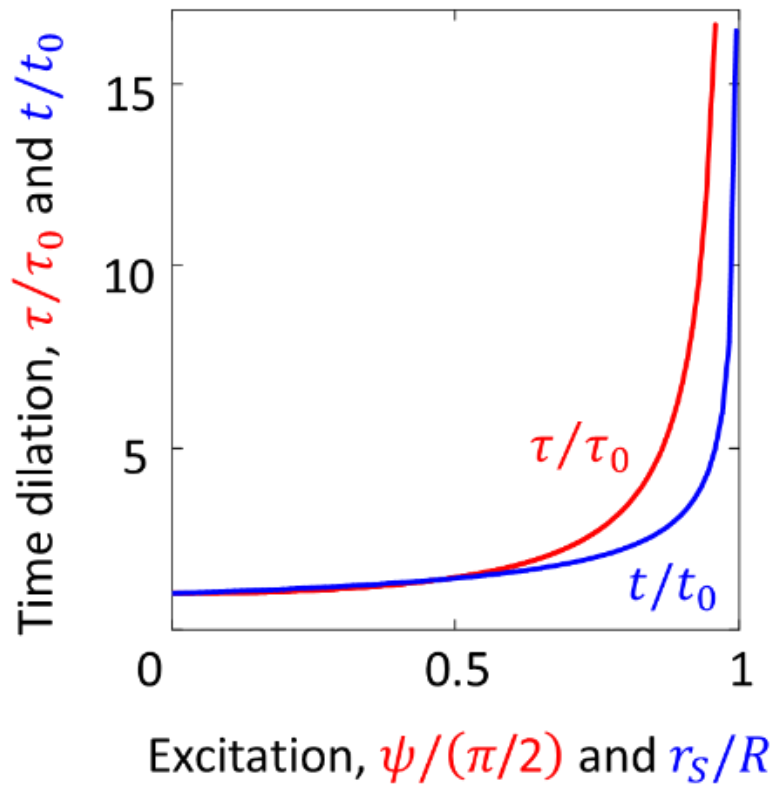

**Figure 4.** Comparison between relativistic gravitational time dilation (blue) and time-crystal time dilation (red).

When two oscillators with initially different oscillation periods are synchronized, each presses the other to operate in the same mode. When a network comprises many of those, their competition is affected by positive feedback: the more oscillators there are in a given mode, the greater their pressure on others to join them. Let us see how this changes the probability distribution $w_{\mathcal{E}}$ given by formula (5) (see also [15]).

Suppose that the probability that an oscillator is in state $\mathcal{E}$ increases linearly with the number of oscillators already in that state, and $k_+$ is the proportionality factor. Then, the synchronization-altered probability $w_+$ can be written as

$$w_+ = w_{\mathcal{E}}(1 + k_+ w_+), \tag{12}$$

where $w_\mathcal{E}$ is the probability before synchronization. After separation of variables

$$w_+ = \frac{w_\mathcal{E}}{1 - k_+ w_\mathcal{E}}, \tag{13}$$

Using expression (5) for the probability $w_\mathcal{E}$ in equation (13), we obtain

$$w_+ \propto \frac{1}{c_+ exp\left(\frac{\mathcal{E}}{\theta}\right) - 1}, \tag{14}$$

where $c_+$ is a coefficient depending on $k_+$. The distribution (14) has a form similar to that of the Bose–Einstein distribution.

Amplitude quenching [16] is another type of synchronization in which the amplitude of the synchronized oscillators is reduced to zero. The affected oscillators group in the zero-amplitude state. Technically, they remove each other from the network. To see how amplitude quenching affects the statistical distribution, we can repeat the previous reasoning with some small alterations in equations (12)-(14): to distinguish between two synchronization scenarios, we use the subindex "$-$" instead of "$+$" and $\mp k_-$ instead of $\pm k_+$. Then, the probability, modified by amplitude quenching, is

$$w_- = w_\mathcal{E}(1 - k_- w_-), \tag{15}$$

which leads to a distribution in a form similar to the Fermi-Dirac distribution:

$$w_- \propto \frac{1}{c_- exp\left(\frac{\mathcal{E}}{\theta}\right) + 1}, \tag{16}$$

where $c_-$ is a coefficient depending on $k_-$.

Unlike in quantum mechanics, quantum-like distributions (14) and (16) do not require an a priori assumption that the oscillators are indistinguishable from each other. The necessary degree of similarity emerges after synchronization.

## 4. Time-crystal phases with topologically different attractors. Spinorial states and diagrams.

Under certain conditions, oscillators experience period doubling. Period doubling is a temporal phase transition accompanied by a symmetry break between oscillator cycles. Theoretically, an oscillator can experience an infinite number of period doublings, a sequence that eventually drives it into chaos. After the first period doubling, the symmetry between even and odd cycles is broken. After the second period-doubling, this happens between the 1st and 3rd cycles and between the 2nd and 4th cycles in each period of four cycles. The period doubling cascade follows certain patterns known as the Feigenbaum universality. However, in practice, chaotic dynamics appear after just a few period doublings.

The states that emerged as a result of period doubling differ from the original in many properties but are perhaps most pronounced in the topology of their attractors (Figure 5). Taking this into account, we call the corresponding phases topological and distinguish them by the attractor index $\mathcal{W}$ (the number of attractor loops/cycles in a complete oscillation period).

We call the oscillator and its topological phase regular if $\mathcal{W} = 1$. Otherwise, we call them spinorial. The name reflects the fact that for $\mathcal{W} \geq 2$, the oscillator can return to any previous state only after a few revolutions around its attractor.

In spinorial phases, each loop/cycle represents a separate spinorial state. Spinorial states are dynamic and replace each other in a certain order. In Figure 5, we conventionally marked different spinorial states with different colors. In the $\mathcal{W} = 2$ spinorial phase, the color sequence alternates between the $black$ and $purple$ cycles: ... $\rightarrow black \rightarrow purle \rightarrow black \rightarrow \cdots$, and in the $\mathcal{W} = 4$ spinorial phase, the color sequence is ... $\rightarrow$ $black \rightarrow red \rightarrow green \rightarrow blue \rightarrow black \rightarrow \cdots$. This color code will be used throughout the paper. The order is fixed and cannot be reversed (due to time irreversibility).

To depict spinorial state dynamics, we use a special type of diagram [15], which we call spinorial clock diagrams (Figure 5, right column). Each color sector in the diagrams covers one spinorial cycle. The diagrams rotate synchronously with the spinorial states (Figure 6(a)). The current state corresponds to the twelve o'clock mark. The direction of rotation is counterclockwise.

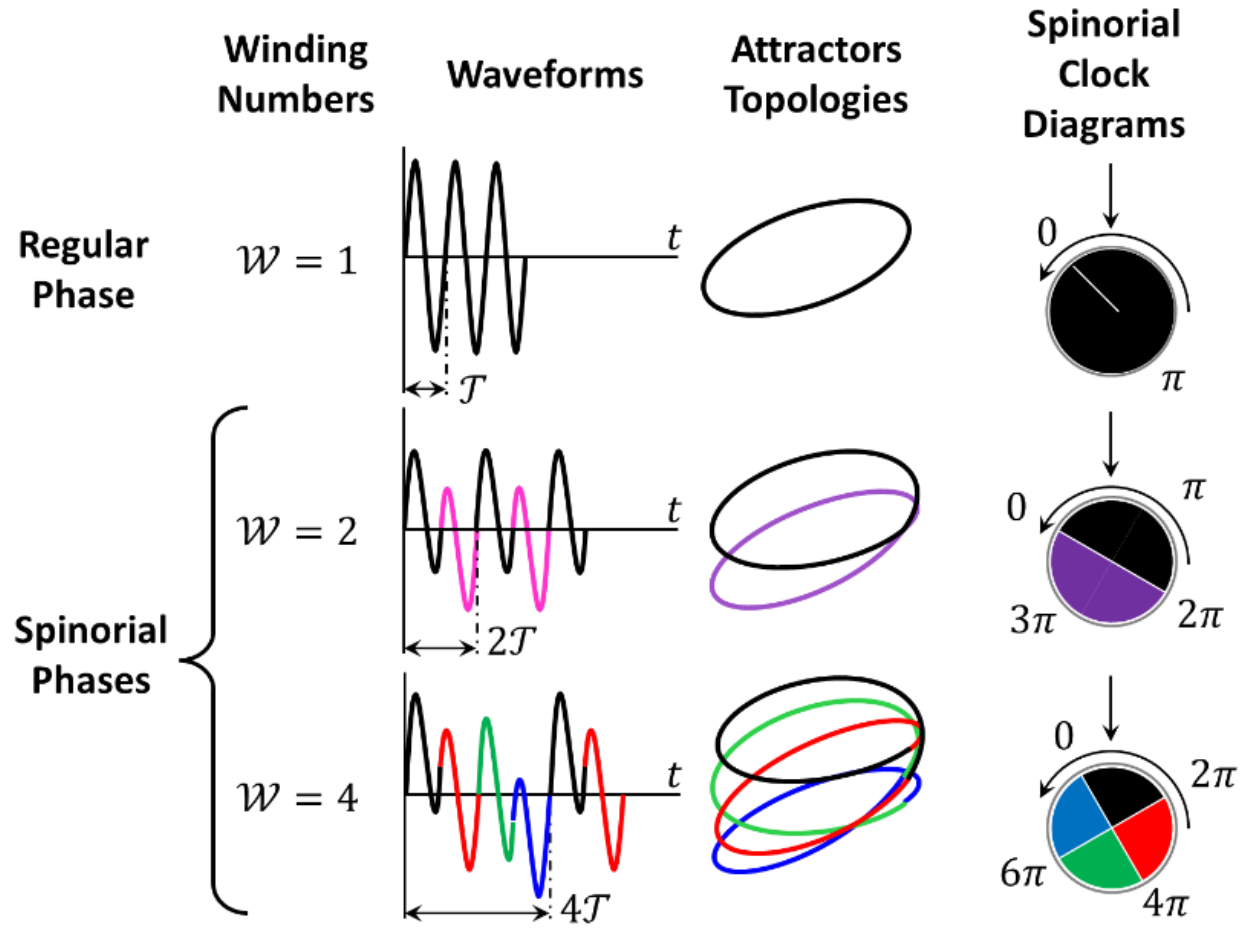

**Figure 5.** Topological phases shown by different types of diagrams. Spinorial states are shown in different colors.

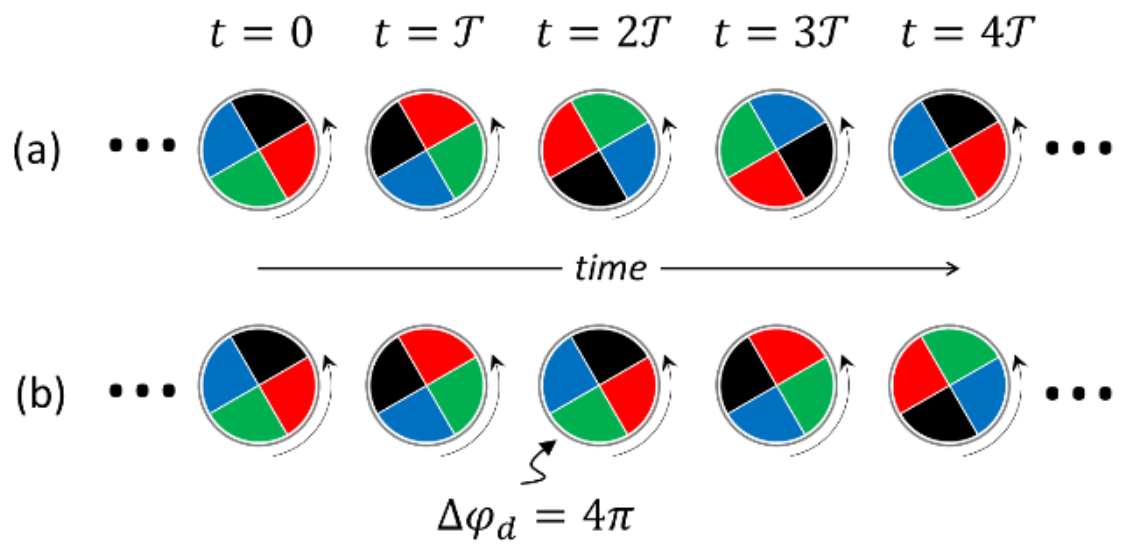

**Figure 6.** (a) Example of $\mathcal{W} = 4$ spinorial clock diagram dynamics, where $\mathcal{T}$ is the duration of one spinorial state; (b) Temporal dislocation $\Delta\varphi_d = 4\pi$.

We call a glitch in the spinorial state sequence that creates a phase shift $\Delta\varphi \neq 0$ for all subsequent oscillations a temporal dislocation and denote as $\Delta\varphi_a$. An example is shown in Figure 6(b). Temporal dislocations strongly affect the properties of links between synchronized spinorial oscillators. This will be addressed in the next section.

Period doublings significantly change the oscillator dynamics, statistical distributions, asymptotic stability, elastic forces, and other properties. As an example, Figure 7 schematically shows how $\mathcal{E}$-profiles change when the oscillator is transformed from one topological phase to another.

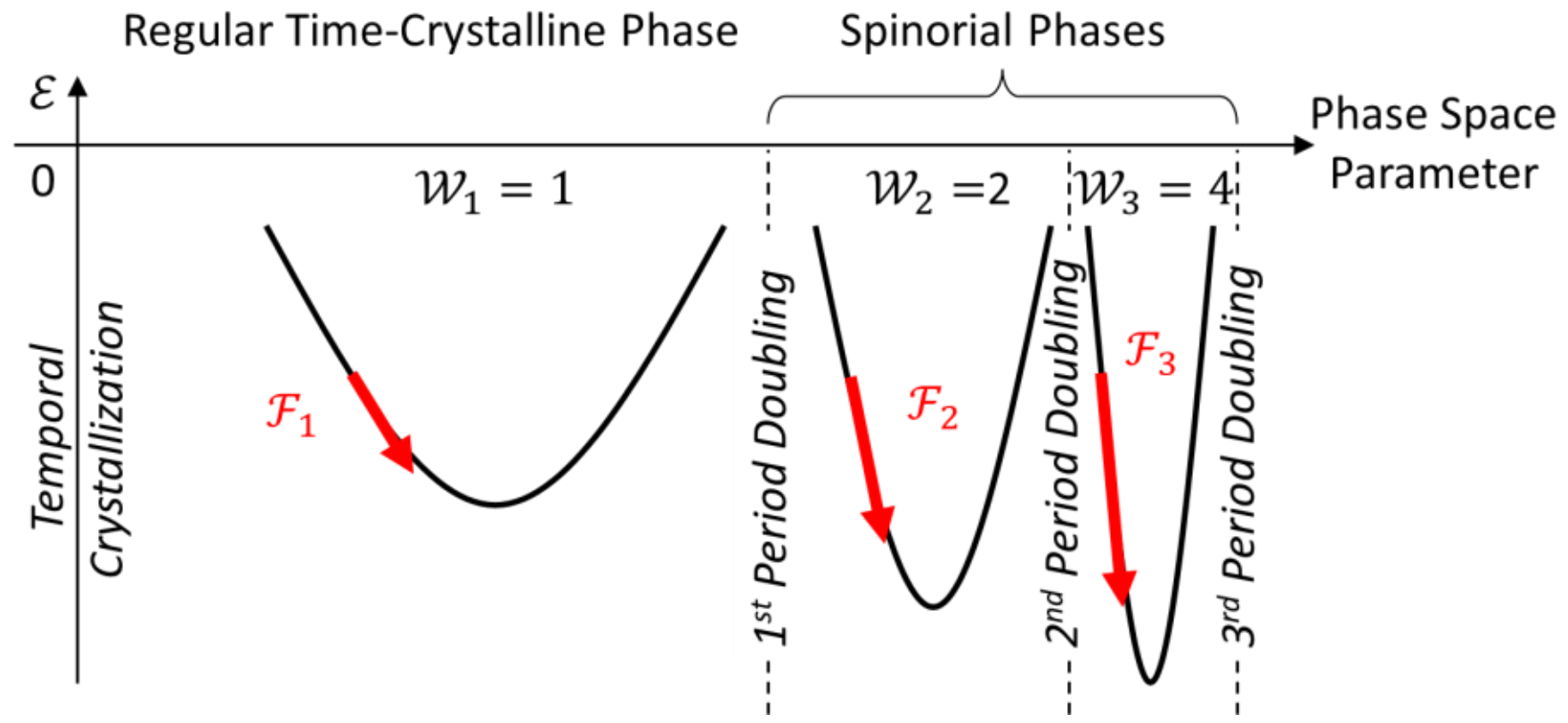

**Figure 7.** $\mathcal{E}$-profiles in different topological phases.

## 5. Link flavors

Spinorial oscillators can be synchronized in the same spinorial states (such as $purple$-$purple$; Figure 8(a) bottom) or different states (such as $purple$-$black$; Figure 8(a) top). The properties of the $purple$-$purple$ links are significantly different from the properties of the $purple$-$black$ links, and we consider them different dynamical systems. However, for links with an attractor index $\mathcal{W} = 4$, the same link can be associated with different colors. For example, a link that is currently $red$-$black$ (Figure 9) will be $green$-$red$ in the next cycle, $blue$-$green$, and so on. However, this link will never turn $red$-$blue$—a combination belonging to a different type of links. The spinorial clock diagrams presented in the previous section help to avoid confusion.

We distinguish links as different if they do not have the same color combinations, regardless of which cycles the comparison is made in. We assign to different links different flavors and flavor numbers $\zeta$. The latter is the phase shift between the coupled oscillators expressed as the number of cycles $\zeta = 0, 1, \ldots \mathcal{W} - 1$, where the count starts from the monochrome links ($\zeta = 0$). Examples are shown in Figures 8 and 9.

Significant differences between links of different flavors are illustrated in Figures 8 and 9. The results were obtained using the iterated map (7) modified by adding subharmonic terms that appear after period doublings. The contributions of the second subharmonic in the iterated map (17) written for $\mathcal{W} = 2$ links are taken into account by the weight coefficient $C_2$, and the contributions of the second and fourth subharmonic in the iterated map (18) written for $\mathcal{W} = 4$ links are taken into account by the weight coefficients $C_2$ and $C_4$.

$$\Delta\varphi_{n+1} = \Delta\varphi_n + \Delta\omega\mathcal{T} - \frac{\kappa}{1+C_2}\left[sin\Delta\varphi_n + C_2\sin\left(\frac{\Delta\varphi_n}{2}\right)\right] \tag{17}$$

$$\Delta\varphi_{n+1} = \Delta\varphi_n + \Delta\omega\mathcal{T} - \frac{\kappa}{1+C_2+C_4}\left[sin\Delta\varphi_n + C_2\sin\left(\frac{\Delta\varphi_n}{2}\right) + C_4\sin\left(\frac{\Delta\varphi_n}{4}\right)\right] \tag{18}$$

The evolutionary trajectories of links with the same parameters $\Delta\omega$, $\kappa$, and weight coefficients but with different flavors $\zeta$ are shown in Figure 8(a). These trajectories evolve to different attractors (fixed points) and occupy different attractor basins. More importantly, they have different relaxation times $\tau$ (Figure 8(b)) and hence stability. As shown in Figure 9, the links with different flavors differ in terms of the parameters $\mathcal{E}$ and $\lambda$, their profiles, ground states, and elastic forces.

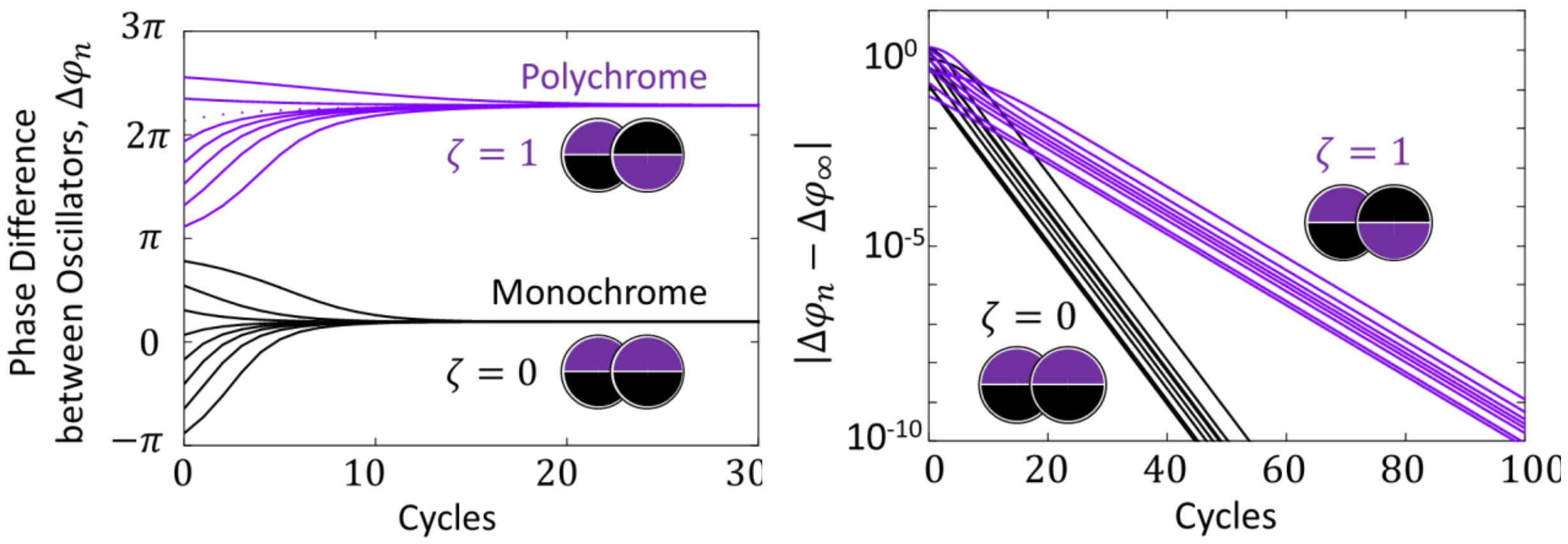

**a**) **(b)**

**Figure 8.** $\mathcal{W} = 2$ bonds: (**a**) evolution trajectories; (**b**) convergence diagram.
$\Delta\omega\mathcal{T} = 0.25;\ \kappa = 0.5;\ C_2 = 0.3.$

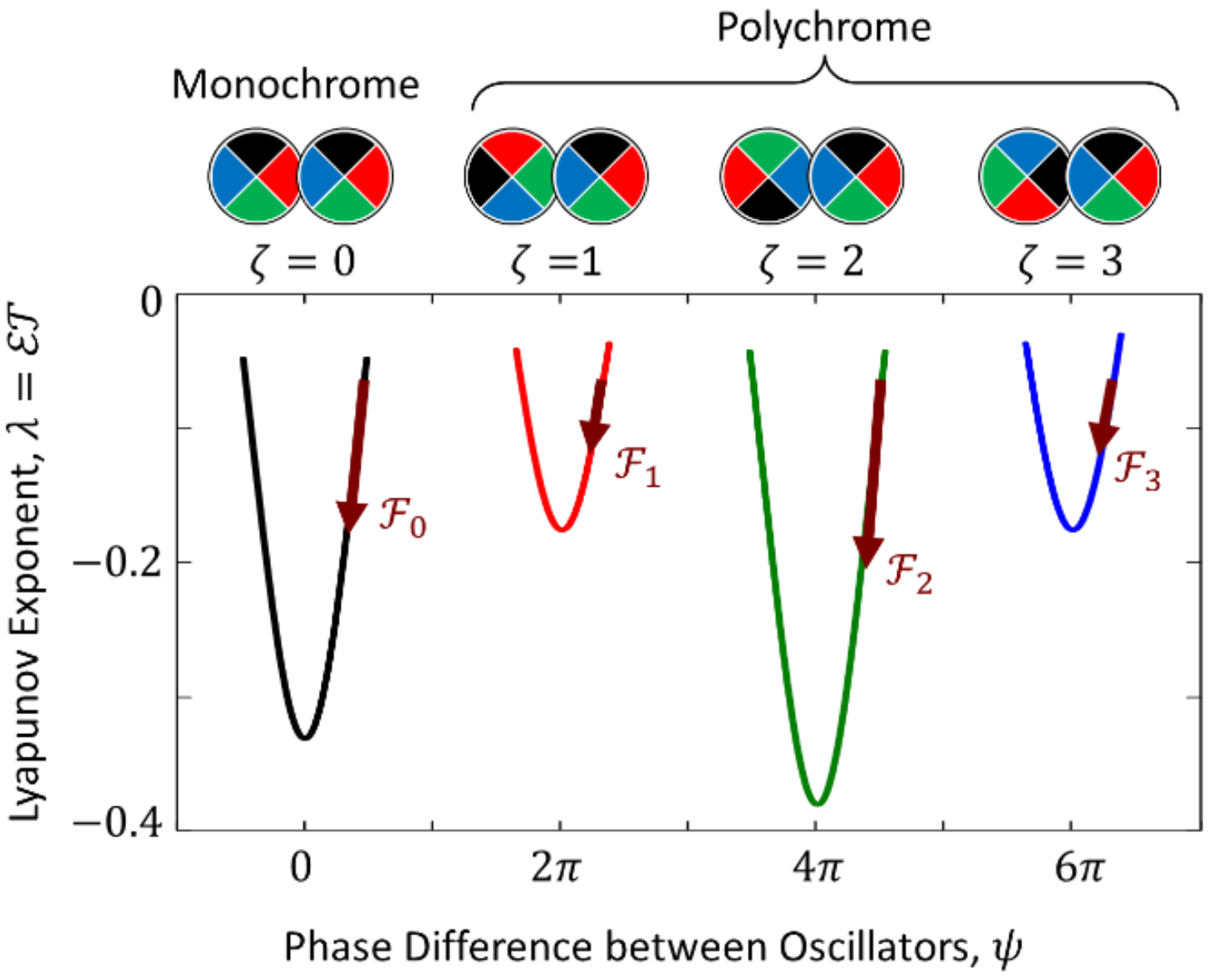

**Figure 9.** $\lambda$-profiles for $\mathcal{W} = 4$ bonds; $C_2 = 0.6;\ C_4 = -0.3$.

In synchronized spinorial networks, phase difference fields arise from the same premises as in networks consisting of regular oscillators. However, due to the variety of spinorial states and flavors of links, the structure of the $\psi$-fields becomes more complex. The latter includes changes in the symmetry of oscillations, the topology of the attractor and the number of degrees of freedom.

## 6. Model of vacuum self-organization

We have briefly reviewed the basic properties of TCs and their networks. Starting with this section, we explore the properties of a hypothetical medium that we associate with vacuum. Below, we assume that the medium consists of dust, the particles of which interact nonlinearly with each other. Dust is controlled by two nonlinear, mutually antagonistic forces, which we call self-attraction and self-repulsion. The pair can be attributed to gravity and self-diffusion or dark energy forces (antigravity). However, for our analysis, the exact nature of these forces is not critical. It is sufficient that they change monotonically with the dust density and act against each other. The medium is far from thermodynamic equilibrium and self-organizes in the form of vortices reminiscent of Bénard cells [23]. The latter belongs to the family of TCs and can undergo period doublings [24].

We do not dwell on the details of the vortex origin. Briefly, this can happen in the following way. Due to self-attraction, randomly formed dust clots create local dust flows directed toward the clot centers. An increase in the clot density increases dust attraction and centripetal flow, which further increases the clot density. This positive feedback would lead to gravitational collapse if not for centrifugal flows due to self-repulsion, which also increases with increasing clot density. Mutually penetrating radial flows in opposite directions are unstable. Flow fluctuations in transverse directions, which are also subject to amplification by the medium, cause spatial separation of radial flows and, ultimately, the formation of circular flows (vortices). The formation of vortices is a phase transition process, which we call temporal crystallization.

The dynamics of a vortex in a turbulent medium are essentially stochastic, and we formalize its evolution using probabilistic variables. We define $p$ as the probability that the radial flows in the vortex are predominantly centripetal and $(1-p)$ as the probability that they are predominantly centrifugal. We assume that with small increments of time $\Delta t$, the probability $p$ changes as

$$p(t+\Delta t) = Ap(t)\big(1-p(t)\big), \tag{19}$$

where $A > 1$ is the medium amplification factor. Below, in addition to $p$ and $A$, we use parameters $\chi$ and $B$, which we call the convergence parameter and feedback control parameter, respectively, and define as

$$\chi = p - (1-p), \tag{20}$$

$$B = \frac{A}{2} - 1. \tag{21}$$

Taking into account some similarity with Gauss's law, the parameter $\chi$ plays the role of the vortex charge, taken with a negative sign.

We describe vortex evolution using an iterated map formalism [14]. Discrete time greatly simplifies equations and numerical analysis. This approach frees the model from cumbersome details imposed by the continuity of trajectories. The iterated map, which replaces equation (19) in terms of the variables $\chi$ and $B$, has the form

$$\chi_{n+1} = \mathcal{L}(\chi_n, B), \tag{22}$$

where sampling occurs at times $t_n$ and $t_{n+1}$ separated by the time intervals equal to the effective vortex rotation period $\mathcal{T}$, and the iteration function $\mathcal{L}(\chi, B)$ is quadratic in the variable $\chi$:

$$\mathcal{L}(\chi, B) = B - (1 + B)\chi^2. \tag{23}$$

We use the latter property to add to vortices their anti-matter counterparts. The latter appears as matter vortices with the reversed directions of flows; hence, the parameter $\chi \to -\chi$. We associate the inversion of flows with a reversal of the forces of self-attraction and self-repulsion and not with the inversion of the arrow of time. With this interpretation, during the transition from matter to anti-matter, attractors remain attractors and do not turn into their opposite (repellers), which would happen in the case of time inversion.

The iterated map for antimatter is

$$\chi_{n+1} = -\mathcal{L}(\chi_n, B). \tag{24}$$

For brevity, the evolution maps of matter and antimatter vortices are combined into one equation:

$$\chi^{\pm}_{n+1} = \pm\left[B - (1 + B)\left(\chi^{\pm}_n\right)^2\right], \tag{25}$$

where "+" and "−" stand for matter and antimatter, respectively. On spinorial clock diagrams, we distinguish matter from antimatter by adding the colored rims black and red, as shown in Figure 10.

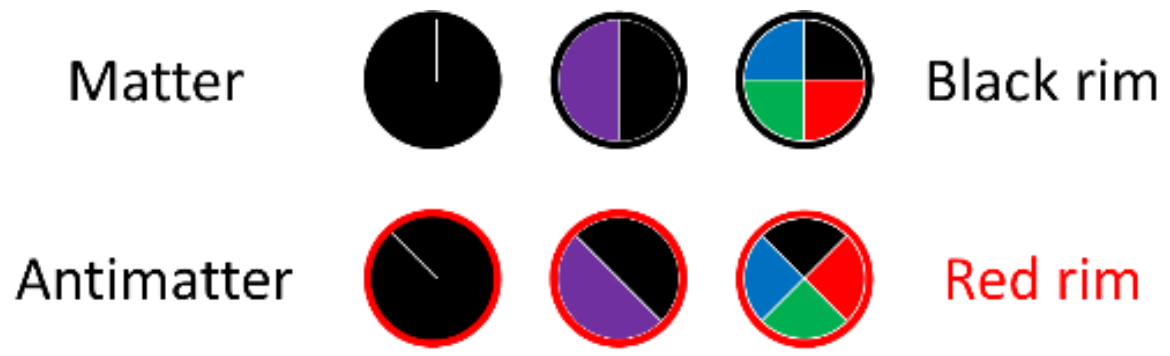

**Figure 10.** Spinorial clock diagrams for matter and antimatter vortices in different vacuum phases.

The quadratic iteration function (23) has a single extremum. This property refers map (25) to the class of systems with Feigenbaum universality [25,26]. Both the qualitative and quantitative results of the model, bearing the imprints of this, depend little on the details of the model formulation, provided that the iteration function remains unimodal.

## 7. Topological phases in time-crystalline vacuum

We simulate vortex evolution using iteration map (25). The minimum number of iterations on each trajectory was chosen so that at the last iteration, the change in the variable $\chi$ did not exceed $10^{-9}$. The $\chi$ values after the last iteration were taken as fixed points. In spinor phases, where $\mathcal{W} > 1$, the trajectories of different spinor ($color$) states converge to different fixed points. Accordingly, for each spinor state, they were calculated separately. The sampling interval was equal to the full vortex period of $\mathcal{W}$ cycles. For each topological phase, the matter and antimatter fixed points form $2\mathcal{W}$-dimensional space, where each direction is marked by its color and + or − superscript, $\chi^{\pm}_{color}$. The relaxation time $\tau$ was determined in the regions near the fixed points using equation (1). The parameters $\mathcal{E}$ and $\lambda$ were determined using equations (2) and (3), respectively. The duration of one cycle in all calculations was assumed to be $\mathcal{T} = 1$. The fixed points simulated for different $B$ values in the three topological phases (denoted by the attractor indices $\mathcal{W}_i$ and $i = 1,2,3$) constitute the bifurcation diagram shown in Figure 11. When comparing the vacuum topological phases with the SM fields, we also refer to these phases as $EM$, $W$ and $S$. Solid and dashed curves depict matter and anti-matter, respectively. The colors of the curves (branches) correspond to the colors of their spinorial states.

One of the most important properties of TCs is their ability to undergo period doubling phase transitions. When applied to vacuum vortices, whose evolution is described by map (25), such phase transitions occur at points $B = B_i$, $B_1 = 0.5$, $B_2 \approx 0.724745$, and $B_3 \approx 0.772045$. For convenience, the $B$-scale in each topological phase is different, and we separate graphs of different topological phases by the period-doubling gaps. $B = B_0 = -0.5$ marks the point at which vacuum temporal crystallization occurs (formation of vortices).

In each topological phase, there exists its own ground state, whose location is designated $B = G_i$, $G_1 = 0$, $G_2 \approx 0.618034$, or $G_3 \approx 0.749281$. In the ground states, all fixed points are $\chi^{\pm}_{color} = 0$.

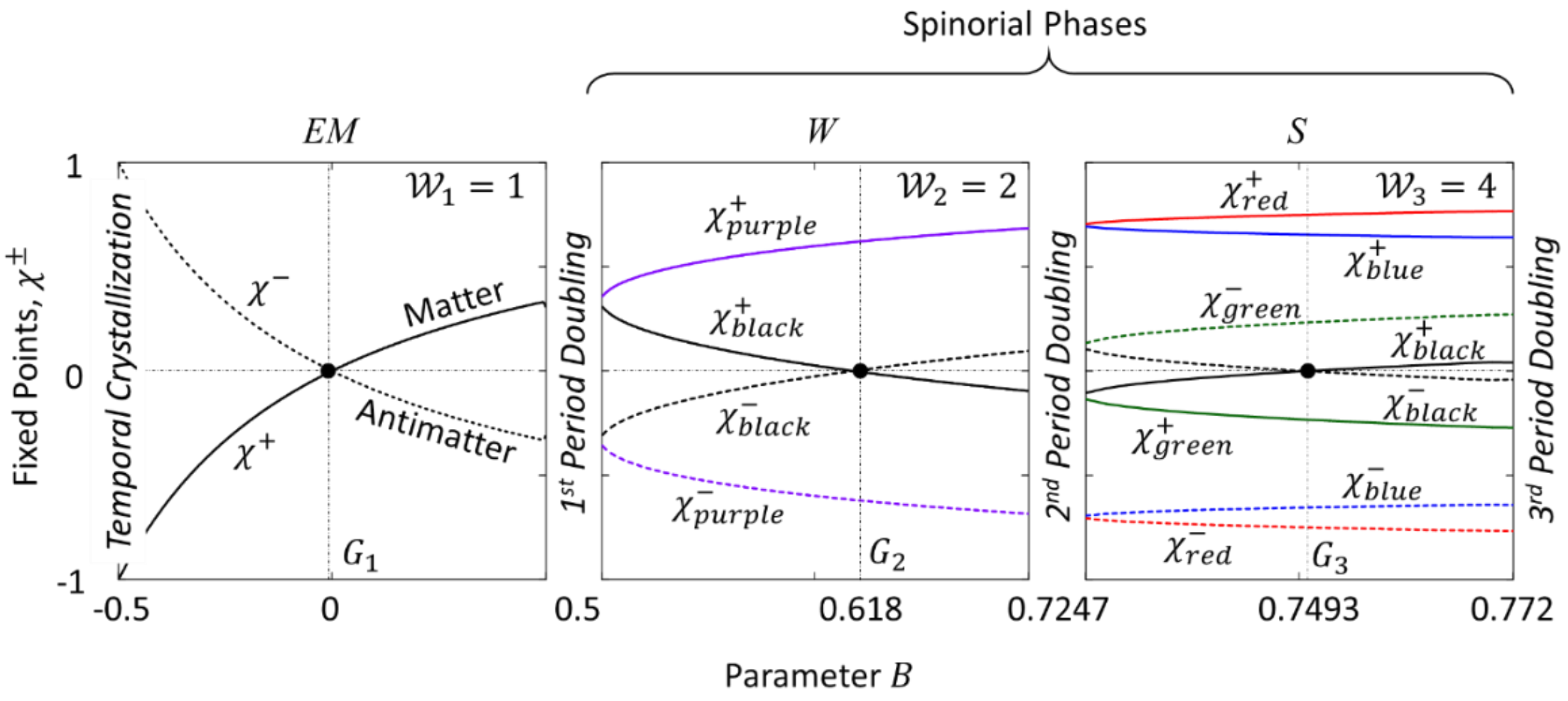

**Figure 11.** Topological-phase bifurcation diagram.

The intervals $B$ occupied by topological phases $EM$, $W$, and $S$ are related to each other as

$$(B_1 - B_0) \div (B_2 - B_1) \div (B_3 - B_2) \approx 1 \div {\delta_F}^{-1} \div {\delta_F}^{-2}, \qquad (26)$$

where $\delta_F \approx 4.6692$ is the Feigenbaum constant.

The function $\mathcal{E}(B)$ in each of the three topological phases (Figure 12) is shaped as a "potential" well, with a minimum at the respective point $B = G_i$. The well walls are logarithmically shaped and can be described as follows:

$$\mathcal{E}_i(B) = \mathcal{T}^{-1} ln(|G_i - B|) - \mu_i, \qquad (27)$$

$$\mu_i = \mathcal{T}^{-1} ln(G_i - B_{i-1}) \approx \mathcal{T}^{-1} ln(B_i - G_i) \tag{28}$$

where the parameter $\mu_i$ plays a role similar to that of the chemical potential.

Equations (27) and (28) are approximate. Here, to avoid the use of additional indices, we neglected slight asymmetry between the left and right halves of the wells in the topological phases $W$ and $S$.

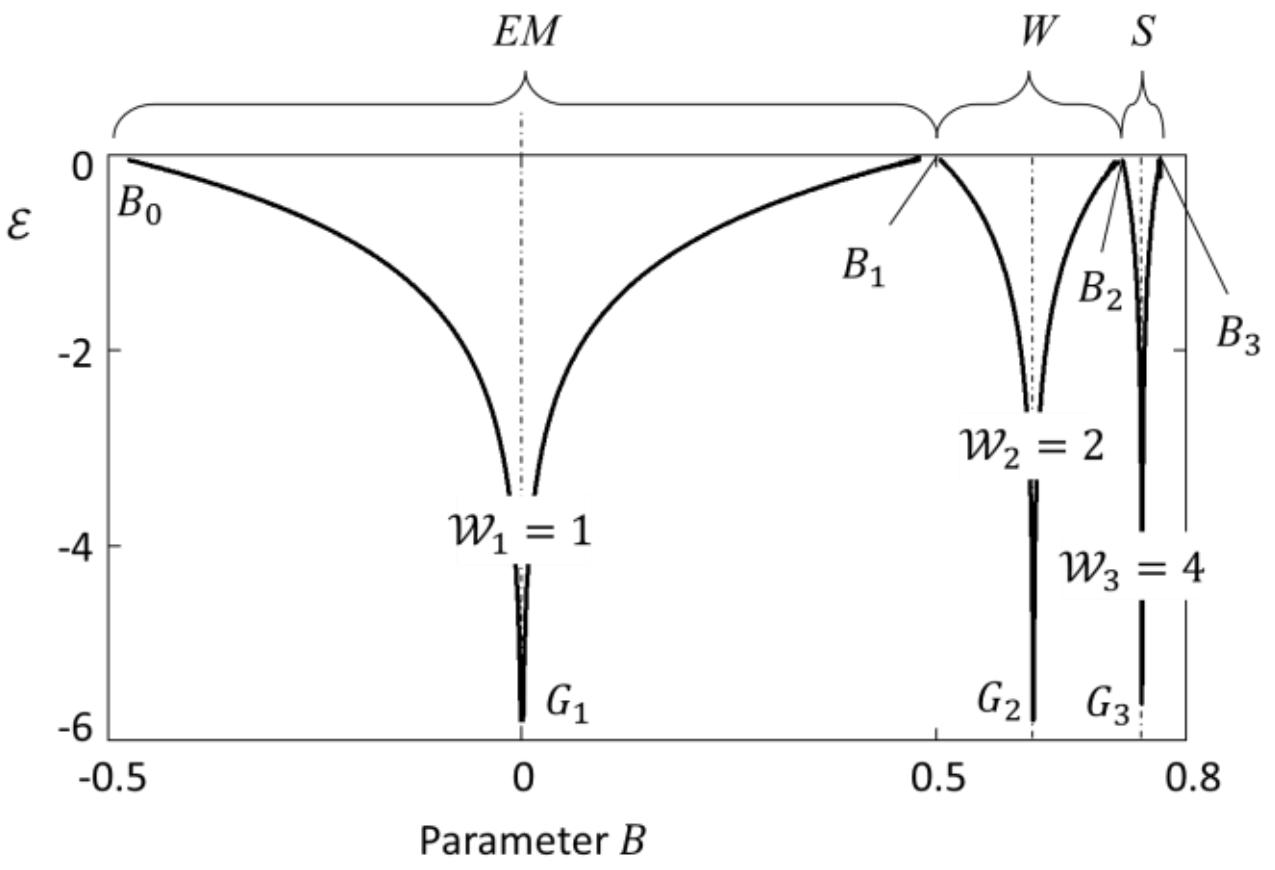

**Figure 12.** $\mathcal{E}(B)$ profile in three vacuum phases.

The distribution of vortices over $B$-space was calculated using equation (6). The probability $w_{\mathcal{E}}$ of finding a vortex in state $\mathcal{E}$ at temperature $\theta$ is

$$w_{\mathcal{E}} = \frac{1}{Z_B} exp\left(-\frac{\mathcal{E}}{\theta}\right), \tag{29}$$

where $Z_B$ is the partition function.

Using equations (27) and (28), the $i$-th topological phase probability $w_{\mathcal{E},i}$ is explicitly expressed as

$$w_{\mathcal{E},i}(B) = \frac{1}{Z_i}\left(-\left|\frac{G_i - B}{G_i - B_i}\right|\right)^{-\theta\mathcal{T}}, \tag{30}$$

where $Z_i$ is the partition function normalizing the probabilities in each topological phase $i$.

As an example, the distribution of vortices at a temperature $\theta = 0.2\mathcal{T}^{-1}$ is shown in Figure 13. At this temperature, the vortices remain on their evolutionary trajectory for an average of only five cycles. As the temperature decreases, the linewidths quickly narrow, and the vortex distribution can be considered discrete.

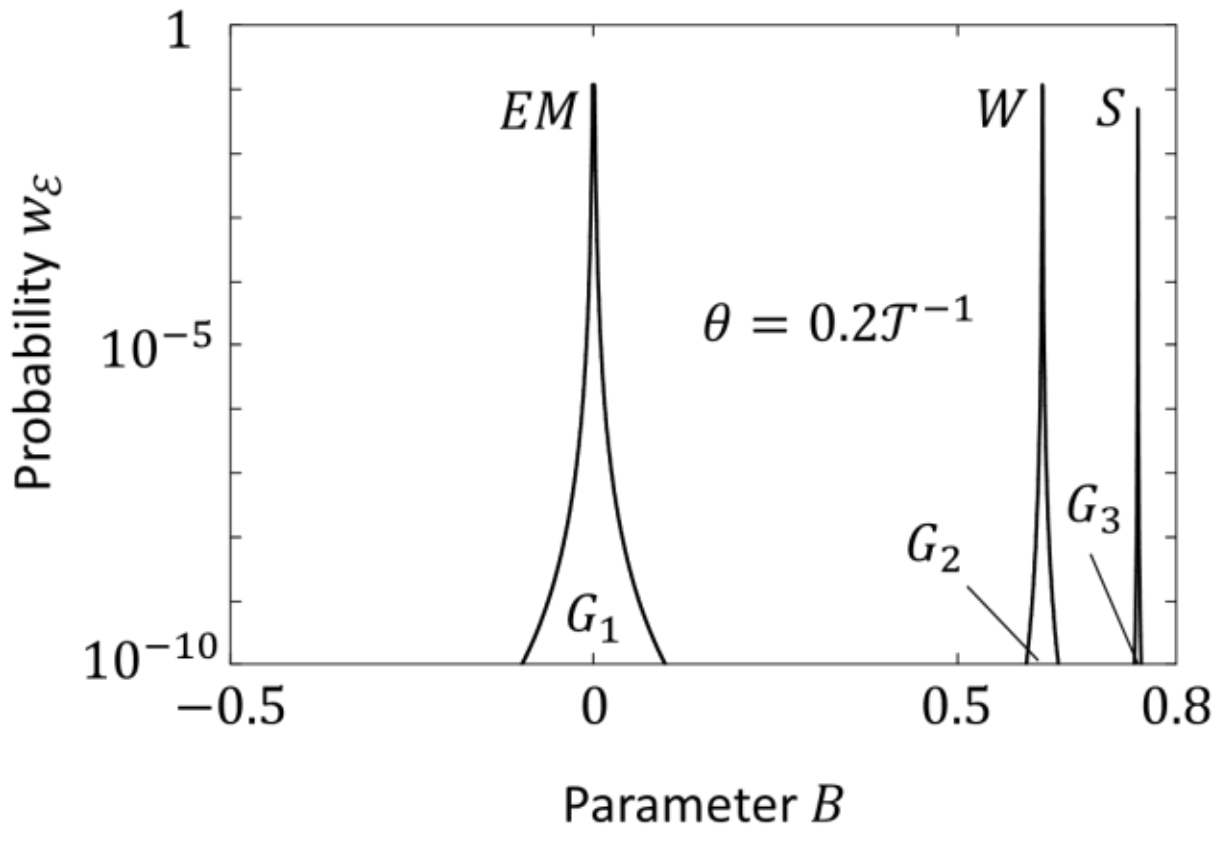

**Figure 13.** Distribution of vacuum vortices in different topological phases.

## 8. Discrete vortex charges and their correspondence to boson charges

The discreteness of the distribution of vortices means that all the parameters dependent on $B$ are also discretized (quantized). With respect to the vortex charges $\chi$, this means that statistically important are only their ground state values $\chi^{\pm}_{color}(G_i)$.

The pattern of vortex charges in topological phases $EM$, $W$ and $S$ resembles the pattern of boson charges in the SM. The correspondence can be seen from Table 1.

**Table 1.** The correspondence chart between vortex charges and charges carried by bosons

| Topological Phase | Vortex Charges | Boson Charges | SM Charge Carriers |
|---|---|---|---|
| $EM$ | $\chi^{\pm}(G_1) = 0$ | 0 | $photon$ |
| $W$ | $-\chi^{+}_{purple}(G_2) > 0$ | $+e$ | $W^+$ |
| | $\chi^{-}_{purple}(G_2) < 0$ | $-e$ | $W^-$ |
| | $\chi^{\pm}_{black}(G_2) = 0$ | 0 | $Z^0$ |
| $S$ | $\chi^{+}_{red}(G_3)$ | $red$ | $gluons$ |
| | $\chi^{-}_{red}(G_3)$ | $anti\text{-}red$ | |
| | $\chi^{+}_{green}(G_3)$ | $green$ | |
| | $\chi^{-}_{green}(G_3)$ | $anti\text{-}green$ | |
| | $\chi^{+}_{blue}(G_3)$ | $blue$ | |
| | $\chi^{-}_{blue}(G_3)$ | $anti\text{-}blue$ | |
| | $\chi^{\pm}_{black}(G_3) = 0$ | 0 | − |
| − | − | 0 | $Higgs$ |

The number of charges in both sets is exactly the same, and they map almost perfectly onto each other. The only exception is the zero-charged $\chi^{\pm}_{black}(G_3)$ in the topological phase $S$, which has no prototype among color-charged gluons, and the zero-charged Higgs boson, which has no prototype in the proposed model.

## 9. Relative strength of elastic forces in different topological phases

The sets of charges are not the only similarities between the time-crystal vacuum and the fields of the SM. In this section, we compare the strength of the forces in each of the models. The subjects for comparison are the elastic forces $\mathcal{F}_i$ acting on the vortices that have left the ground state in different topological phases and the field coupling constants $\alpha_i$ in the SM. Both types of parameters, despite their names, are not constants. Therefore, for each of the models, we compare the results at a fixed level of excitation.

Like in expression (8), we define the elastic force in the topological phase $i$ as

$$\mathcal{F}_i = -d\mathcal{E}_i/dB. \quad (31)$$

As follows from equations (22) and (23), because of the logarithmic shape of walls in $\mathcal{E}$-wells, the ratio of elastic forces $\mathcal{F}_i(\mathcal{E})/\mathcal{F}_j(\mathcal{E})$ in any two different topological phases $i$ and $j$ is a constant and, taking into account the Feigenbaum universality, according to equation (26), is equal to

$$\frac{\mathcal{F}_i(\mathcal{E})}{\mathcal{F}_j(\mathcal{E})} = \frac{B_j - G_j}{B_i - G_i} \approx {\delta_F}^{i-j}, \quad (32)$$

where $\delta_F$ is the Feigenbaum constant and $\mathcal{E}$ is, although chosen arbitrarily, the same in both phases. In (32), we neglected the weak asymmetry of the $W$ and $S$ wells, which slightly altered the ratios when comparing the left-side wall forces with the right-side wall forces. After rounding to two significant figures, the elastic forces in the $EM$, $W$ and $S$ wells relate to each other as

$$\mathcal{F}_{EM} \div \mathcal{F}_W \div \mathcal{F}_S \approx 1 \div \delta_F \div {\delta_F}^2 \approx 1 \div 4.7 \div 22. \quad (33)$$

These ratios are in good agreement with the ratios of the coupling constants $\alpha$, $\alpha_W$, and $\alpha_S$ of electromagnetic, weak, and strong forces, which, at the energies of vector boson masses $m_W$ and $m_Z$, relate to each other as follows (after rounding to two significant figures):

$$\alpha \div \alpha_W \div \alpha_S \approx 1 \div 4.5 \div 15. \quad (34)$$

In (34), the estimates are based on empirical values recommended by the Particle Data Group [27].

## 10. Transfer mechanism of spinorial states

In networks, oscillators can be connected either directly or through other oscillators, through both serial chains and parallel paths. The degree to which any two oscillators are correlated depends on the states of all oscillators in the network, and in some situations, an oscillator may correlate with a distant oscillator better than with its neighbors [22]. Since the stability of the links depends on the phase correlation between the respective oscillators, this can lead to strange consequences. Let us consider two phenomena associated with a strong phase correlation between vortices distant from each other.

A spinorial state as an excitation can move through the network, jumping from one vortex to another. The probability of where the spinor state ends up after a jump depends on how strongly the vortex from which the jump originates correlates with the vortex in which it ends. The jump can occur not between nearest neighbors, but between vortices, although distant from each other, but having closer phases, as shown in Figure 14.

In Figure 14, a spinorial state jumps from the initial position I, shown by the red circle in Figure 14(a), to the final position F, shown by the red circle in Figure 14(c). During the transfer (Figure 14(b)), the spinorial state spreads over all "white" vortices, which participate in the transfer. It is rather virtual and has no specific location. The jump resembles quantum tunneling.

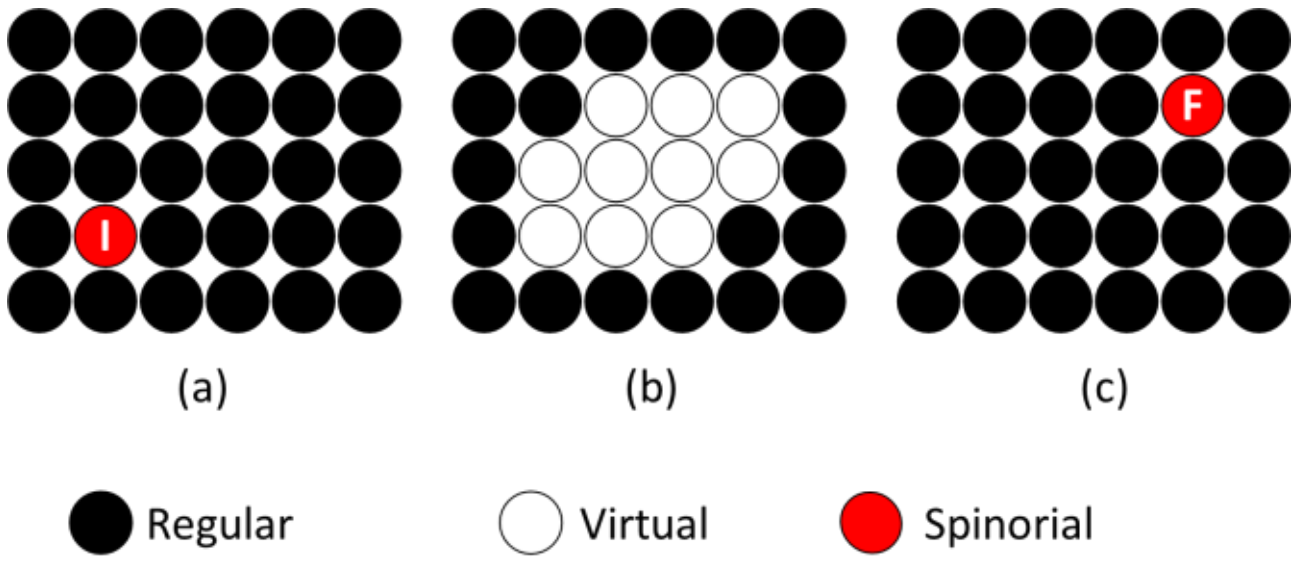

**Figure 14.** Three steps of spinorial mode tunneling from position I to position F: (**a**) before transfer, (**b**) during transfer, and (**c**) after network relaxation.

The excitation of the vortex can be noticeably shorter than its relaxation. In some cases, it can be assumed that all "white" vortices are excited almost simultaneously and that the entire transfer time is as short as the relaxation time of a single vortex. The total time required to transfer the spinor state can then be much shorter than the time required for the waves to propagate through the network between points I and F, apparently violating the principle of relativistic locality.

Similar behavior may occur when two spinorial states are transferred to two positions, $F_1$ and $F_2$, distant from each other and from source I, as shown in Figure 15. As in the previous example, vortices $F_1$ and $F_2$ are strongly correlated with vortex I; thus, they are strongly correlated (entangled) with each other. It is important to emphasize that entanglement occurs not because the spinorial states carry some hidden parameters but because they are transferred to the vortexes with a prior strong correlation. As in the case of "tunneling," "entanglement," in some cases, may apparently "violate" the principle of relativistic locality.

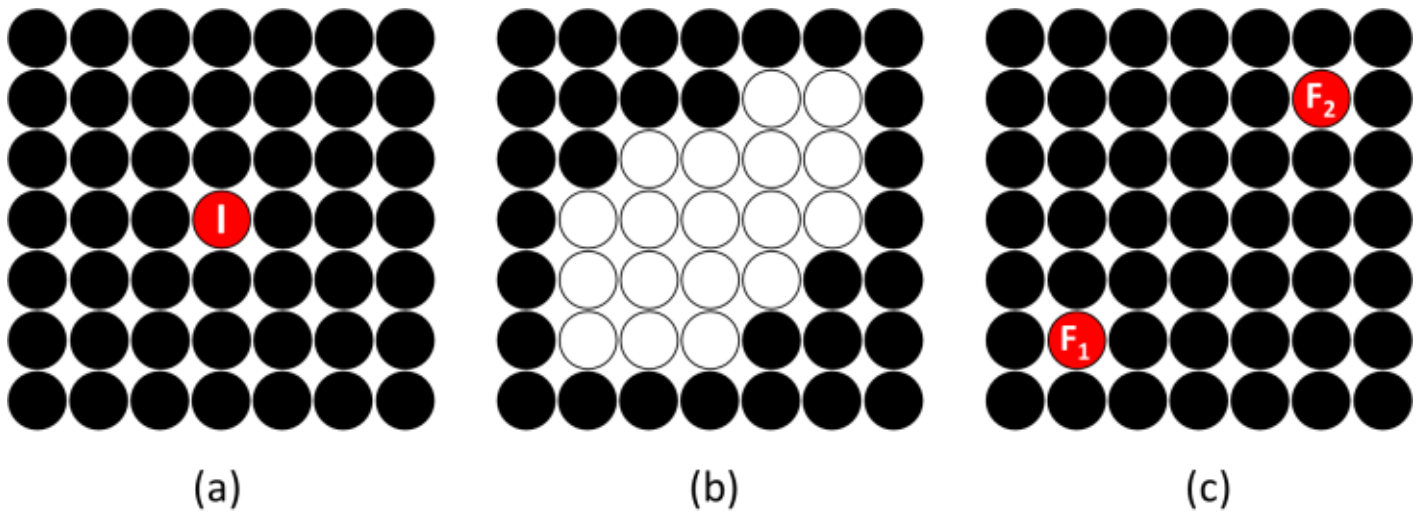

**Figure 15.** Transfer of split spinorial modes from initial position I to final positions $F_1$ and $F_2$: (**a**) before excitation, (**b**) during excitation, and (**c**) after network relaxation.

## 11. Spinons as particle building blocks

As we have discussed, in synchronized TC networks, phase-difference fields emerge as a result of oscillator (vortex) phase entrainment. The elementary carriers of phase differences – the field elements – are the direct links between the vortices. In this paper, we consider three types of fields and links: $EM$ ($\mathcal{W} = 1$), $W$ ($\mathcal{W} = 2$), and $S$ ($\mathcal{W} = 4$). In spinorial networks ($\mathcal{W} > 1$), the links differ by the nature of their vortices (matter or antimatter), by the attractor index $\mathcal{W}$, and also by their synchronization patterns (parameter $\zeta$). We address all of these factors by ascribing different flavors to different links.

The dynamics of the links between vortices, which belong to the same topological phase, have the same attractor topology as its vortices. We call them pure links and differentiate by their attractor winding numbers $\mathcal{W}$. In hybrid links, the vortices belong to different topological phases. In these links, the vortex with the higher attractor index $\mathcal{W}$ influences the dynamics of its counterpart. As a result, the temporal symmetry of the latter is broken, and the topology and attractor index $\mathcal{W}$ of the former are effectively acquired. The influence can extend to the vortices indirectly connected to the first vortex, and an originally pure regular link due to the presence of spinorial vortices in their neighborhood may acquire spinorial dynamics.

Below we will discuss prototypes of TC of elementary particles. For brevity, we will call them particles and use the full name in cases where they can be confused with real particles. We assume that each particle consists of a core and a surrounding cloud. We will need to consider different particle flavors, so it is more convenient to describe particle composition in terms of fields and links. In this regard, the core consists of spinorial links, and the cloud is composed of the hybrid links with spinorial properties induced by their neighbors. The core is the carrier of charge and flavor of the particle. The cloud is mainly responsible for energy (mass).

If a link consists of one matter vortex and one antimatter vortex, and both operate as a monochrome pair $\zeta = 0$, the vortices are subjects to mutual annihilation. If both vortices belong to matter or both belong to antimatter, then again, a polychrome synchronization ($\zeta \neq 0$) is more stable than the monochrome one, because it is less perturbative for the environment due to the smoothing the amplitude of charge oscillations. Therefore, as parts of stable particle cores, we consider only polychrome links.

We call a core link with its vortices a spinon (Figure 16). A compact spinon configuration is shown in Figure 16(a). Altogether, we use six spinons, three belonging to the $W$ phase (Figure 16(b)), and three belonging to the S phase (Figure 16(c)).

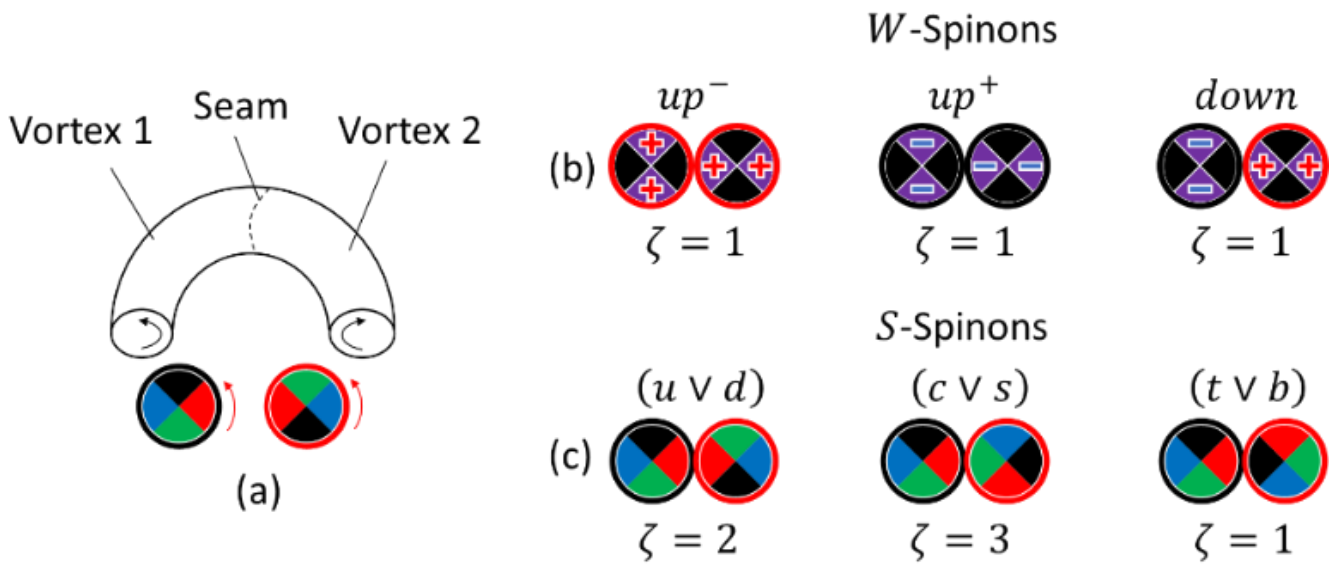

**Figure 16. (a)** Spinon torus (cut in half to show the directions of circular flows); (**b**) $W$-spinon flavors; (**c**) $S$-spinon flavors.

We designated the $W$-spinon flavors $up^+$ (polychrome link with both matter vortices), $up^-$ (polychrome link with both antimatter vortices), and $down$ (polychrome link with one mater vortex and one antimatter vortex). After time averaging, the $down$ spinons are neutral, and $up^{\pm}$ are charged. Due to polychrome synchronization, the charge is carried by one of the vortices during half of the total period and by the other vortex during the other half of the total period. Based on the data given in Table 1, we assign electric charges $\mp e$ to the $up^{\pm}$ spinons. The "+" and "–" superscripts refer to matter and antimatter, respectively (they are opposite to the charge polarity).

$S$-spinons consist of one matter vortex and one antimatter vortex. After time averaging, the $S$-spinons are electrically neutral and color blind. We designated the $S$-spinon flavors of unperturbed spinons as "$u$ or $d$", "$c$ or $s$", and "$t$ or $b$" and used the notation $(u \vee d)$, $(c \vee s)$, and $(t \vee b)$, respectively, where $(\ldots \vee \ldots)$ stands for "or". The names are borrowed from real quarks as combinations of the up type and down type quark flavors. When an $S$-spinon and $W$-spinon are coupled to each other, they exchange circular and radial flows that lead to their hybridization. As a result, the $S$-spinon acquires part of the $W$-spinon charge. The couple, as a whole, still carries the integer $\mp e$ charge, but the charges carried by both spinons are fractional. On the other hand, the symmetry of the dynamics of the $W$-spinon breaks, and its winding number increases from $\mathcal{W} = 2$ to $\mathcal{W} = 4$. As a result, both spinons have fractional charges and the same attractor topology. The symmetry between the up type $S$-spinons and down type $S$-spinons also breaks. If the $S$-spinon is coupled to a charged $up^{\pm}$-spinon, we designate its flavor as $u^{\pm}$, $c^{\pm}$, or $t^{\pm}$, and when it is coupled to a $down$-spinon, we designate its flavor as $d$, $s$, or $b$. This is summarized in Table 2 and is formalized as the direct product (35) of pure spinon flavors.

**Table 2.** Flavors of hybrid spinons

| | $(u \vee d)$ | $(c \vee s)$ | $(t \vee b)$ |
|---|---|---|---|
| $up^{\pm}$ | $u^{\pm}$ | $c^{\pm}$ | $t^{\pm}$ |
| $down$ | $d$ | $s$ | $b$ |

$$\begin{pmatrix} u\,c\,t \\ d\,s\,b \end{pmatrix}_{\pm} = \big((u \vee d), (c \vee s), (t \vee b)\big) \otimes \begin{pmatrix} up^{\pm} \\ down \end{pmatrix}. \tag{35}$$

An example of a flavor change in the hadron environment, from positive $c^{-}$ to neutral $d$, is shown in Figure 17.

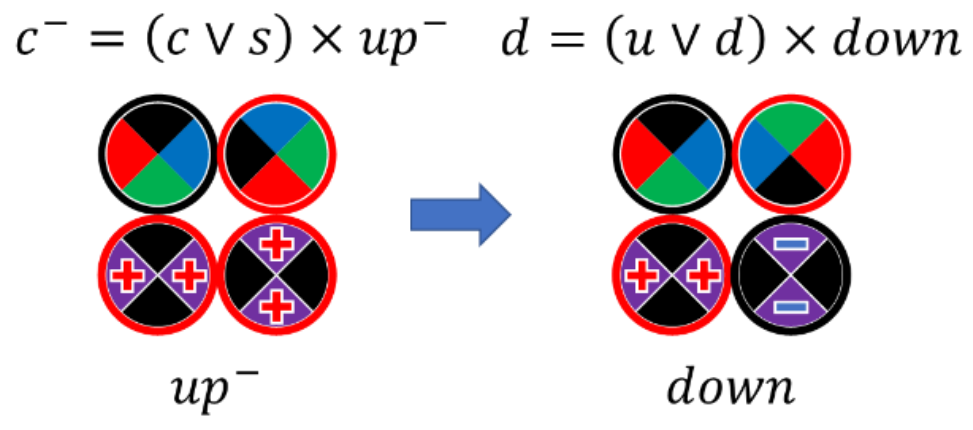

**Figure 17.** Example of a flavor change from $c^{-}$ to $d$.

## 12. Examples of time-crystal particles

In this section, we will give examples of particle cores. Each particle differs from the other in its spinon composition. We did not address their spatial arrangement, which affects the energy of links thus the particle masses.

The lepton cores are shown in Figure 18. The first-generation leptons are represented by a single $W$-spinon (Figure 18(a)). The cores of the second generation and the third generation are added one (Figure 18(b)) and two (Figure 18(c)) neutral $down$-spinons, respectively.

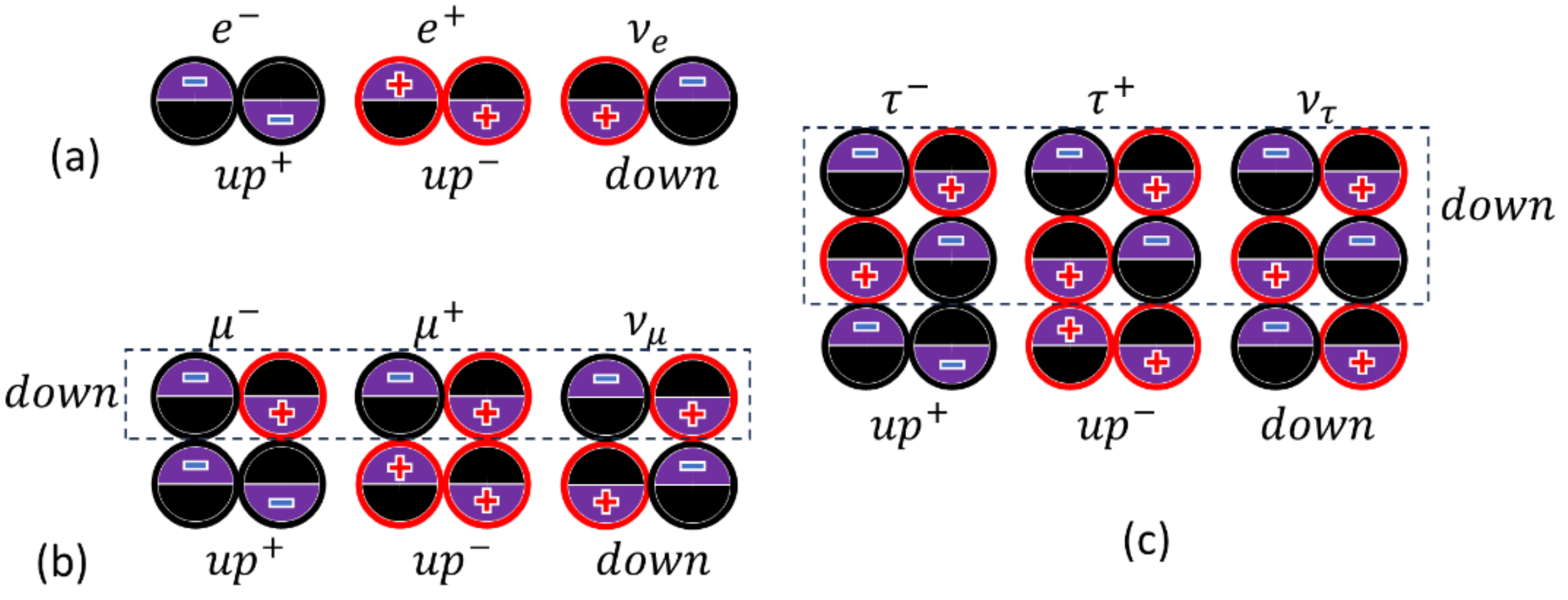

**Figure 18.** Particle cores of three generations of prototype leptons.

The eight mesons shown in Figure 19 mimic the Gel–Mann "eightfold way" real meson set. They consist of one $W$-spinon and one $S$-spinon, both of which are hybridized. Out of three existing $S$-spinons, in the octet, we used only two, $(u \vee d)$ and $(c \vee s)$. The mesons can be seen as an upgrade of the second-generation leptons by replacing one $down$ spinon with one $S$-spinon.

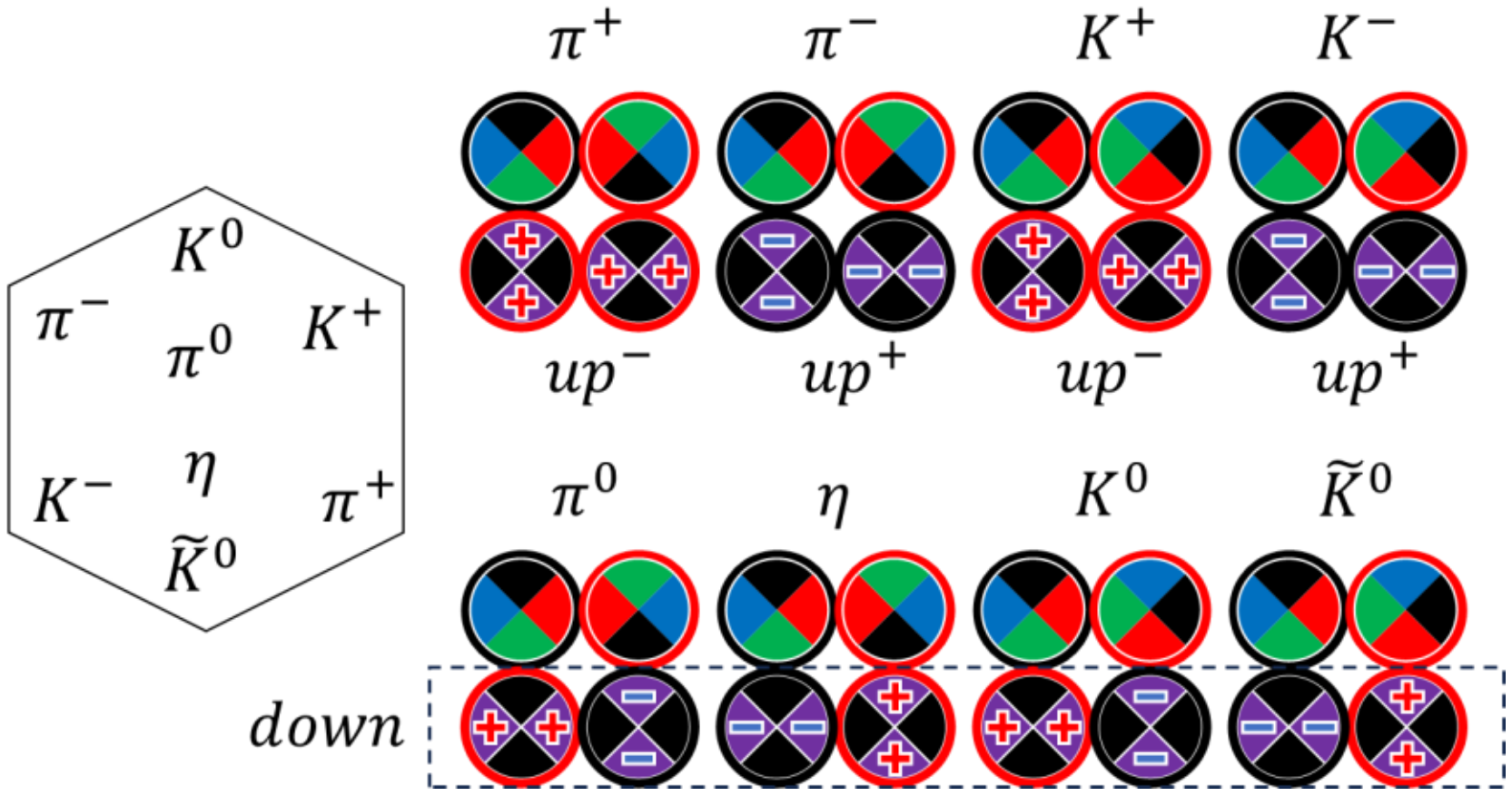

**Figure 19.** Spinon representation of "eightfold way" mesons.

The eight baryons shown in Figure 20 mimic the Gel–Mann "eightfold way" real baryon set. The baryons consist of one $W$-spinon and two $S$-spinons. In this set of baryons, we also used only two types of $S$-spinons, $(u \vee d)$ and $(c \vee s)$. The baryons can be seen as an upgrade of the third-generation leptons where two $down$ spinons are replaced with two $S$-spinons.

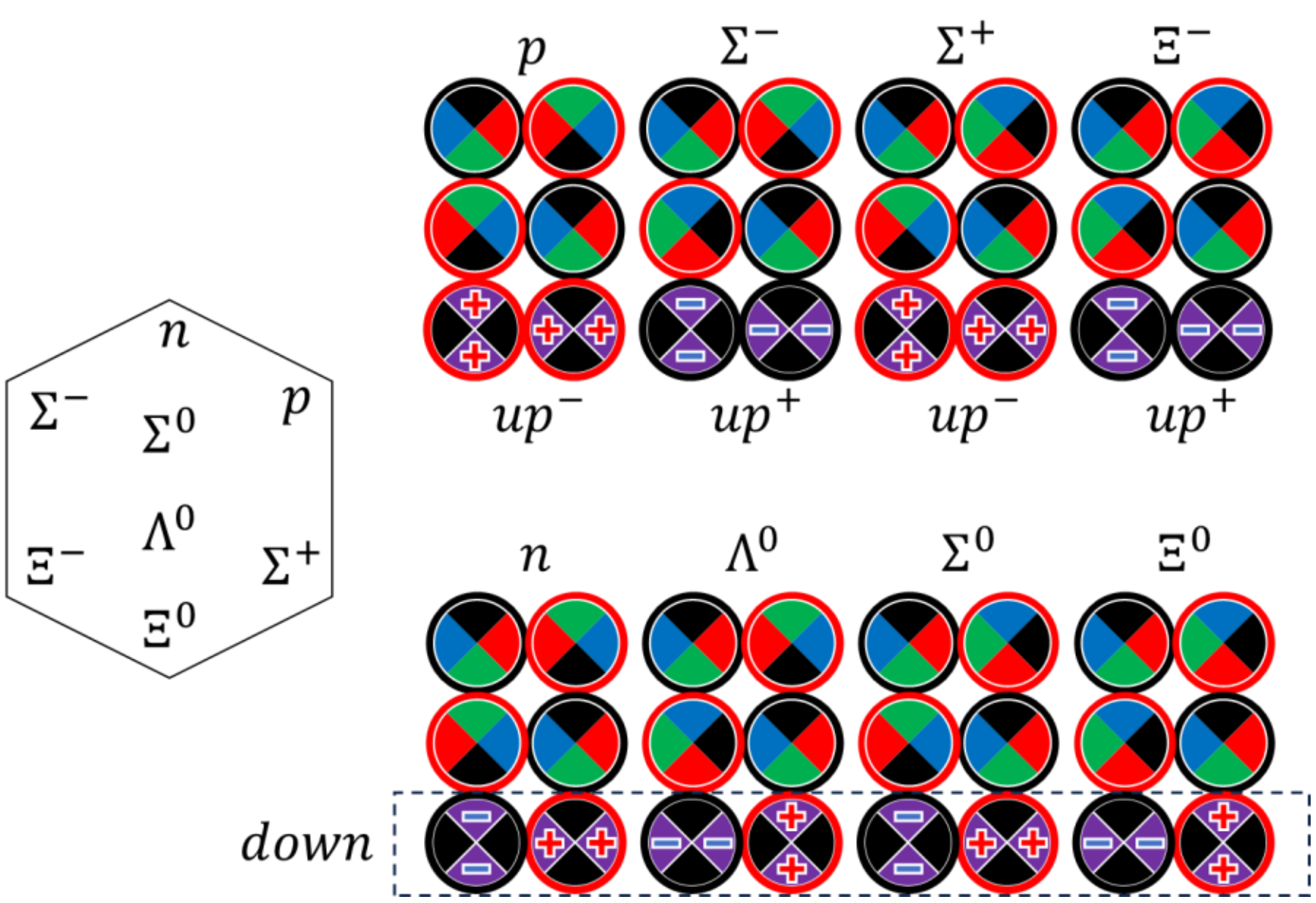

**Figure 20.** Spinon representation of "eightfold way" baryons.

In Figure 20, only particles of matter are shown. The antiparticles can be obtained by replacing $up^{\pm}$ spinons with $up^{\mp}$spinons.

## 13. S-flavor rotations

In this section, we estimate the probabilities of flavor changes among the $S$-spinons. In addition to the three stable polychrome $S$-spinons $(u \vee d)$, $(c \vee s)$ and $(t \vee b)$, we also need the virtual monochrome spinon $(x \vee y)$, which has no counterpart among the quarks and cannot be present in the stable particle states.

A spinon flavor change occurs as a result of a temporal dislocation (abrupt phase shift) $\Delta\varphi_d$ in one of the coupled vortices, while the other is unperturbed or experiences a different phase shift. Without losing generality, we assume that the second vortex is unperturbed. The perturbation acting on the first vortex moves it for a short time $\Delta t$ from the color-reach topological phase $S$ to the color-blind topological phase $EM$. After perturbation, the vortex returns to phase $S$ in the same phase (spinorial state), from which it has departed. The second vortex remains in phase $S$ at all times and maintains its normal pace. We assume that the probability $w$ of a brief visiting the topological phase $EM$ decreases exponentially with increasing time interval$\Delta t$

$$w \propto exp(-\Delta\mu\Delta t), \qquad (36)$$

where $\Delta\mu = \mu_{EM} - \mu_S$ is the difference between the "chemical potentials" in the topological phases $EM$ and $S$.

We also assume that before and after perturbation, the vortex link is in one of its ground states (well bottoms in Figure 9, separated from each other by $2\pi$ intervals); thus, $\Delta t$ is equal to an integer number of cycles:

$$\Delta t_j = j\mathcal{T}, j = 0, 1, 2, 3. \quad (37)$$

where $\mathcal{T}$ is the cycle duration and $j$ is the number of spinorial states skipped by the affected vortex during the stay in phase $EM$. In (37), because of the exponential decay of probability with time, we limit $\Delta t$ to one full vortex period (four cycles). Longer perturbations can be considered small corrections to the results shown below.

From equations (28), (36), and (37), and taking into account relation (26),

$$\Delta\mu \approx \frac{1}{\mathcal{T}} ln\left(\frac{G_S - B_S}{G_{EM} - B_{EM}}\right) \approx \frac{1}{\mathcal{T}} ln\left({\delta_F}^{-2}\right), \quad (38)$$

and the probability $w_j$ of the events decreases with the number of skipped cycles by the first vortex as

$$w_j \propto {\delta_F}^{-2j}. \quad (39)$$

The flavor rotations according to the above scenario are shown in Figure 21 by the arch arrows. The direction of flavor rotation depends on which vortex (matter or antimatter) skips cycles. Figure 21(a) shows the flavor rotation sequence when a matter vortex is affected by temporal dislocation, and Figure 21(b) shows the opposite case. The initial states of flavor rotations (starting arrow positions) are marked with thick points.

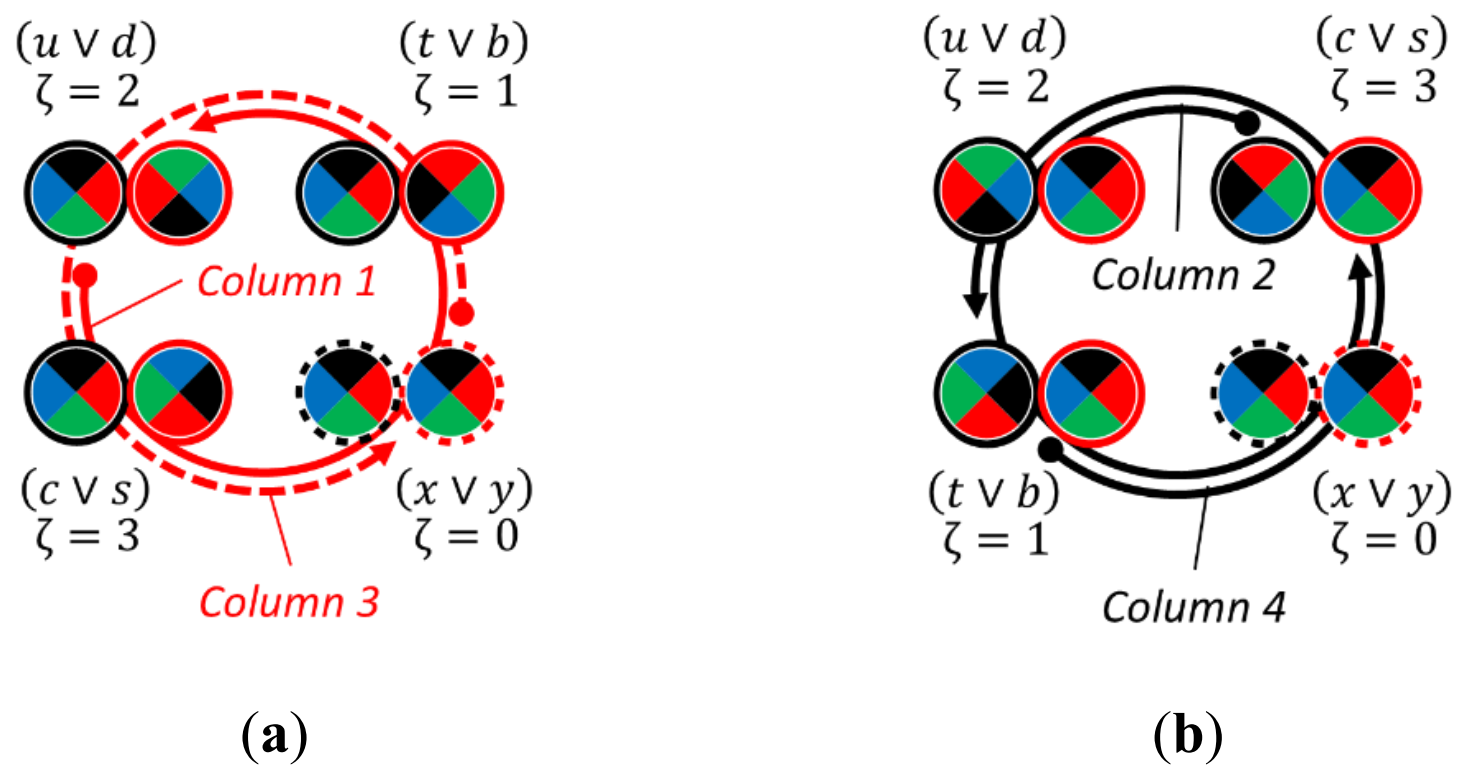

**Figure 21.** Flavor rotation diagrams. The perturbation-affected vortex is (**a**) matter and (**b**) antimatter.

The relative probabilities of transitions from one flavor to another caused by flavor rotation are given in Table 3. Because we did not account for spinon hybridization, which affects the vortex "chemical potential," the matrix is symmetric: the rows and columns can be swapped. The odd rows (columns) start with flavors $(u \vee d)$ and $(x \vee y)$ and reflect the events shown in Figure 21(a), and the even rows (columns) start with flavors $(c \vee s)$ and $(t \vee b)$ and reflect the events shown in Figure 21(b). The initial and final spinons should be stable. Therefore, the rows and columns with starting or ending $(x \vee y)$ flavors (shown in red) should be removed from the finalized table. Removing them reduces the matrix to a size of $3 \times 3$. To convert relative probabilities to absolute probabilities, we used unitary conditions $\sum w_j = 1$ for each row (column).

**Table 3.** Probabilities $w_j$ of flavor change in nonhybridized $S$-spinons

| | $(u \vee d)$ | $(c \vee s)$ | $(x \vee y)$ | $(t \vee b)$ |
|---|---|---|---|---|
| $(u \vee d)$ | ${\delta_F}^0$ | ${\delta_F}^{-2}$ | ${\delta_F}^{-4}$ | ${\delta_F}^{-6}$ |
| $(c \vee s)$ | ${\delta_F}^{-2}$ | ${\delta_F}^0$ | ${\delta_F}^{-6}$ | ${\delta_F}^{-4}$ |
| $(x \vee y)$ | ${\delta_F}^{-4}$ | ${\delta_F}^{-6}$ | ${\delta_F}^0$ | ${\delta_F}^{-2}$ |
| $(t \vee b)$ | ${\delta_F}^{-6}$ | ${\delta_F}^{-4}$ | ${\delta_F}^{-2}$ | ${\delta_F}^0$ |

Because of the similarity between spinons and quarks, it would be interesting to compare the matrix of spinon flavor changes with the CKM matrix (Cabibbo-Kobayashi-Maskawa) [28], which tabulates empirical values of the quark mixing amplitudes (Table 4(b)) and has the same size. To do this, we replaced probabilities with their amplitudes $A_j = {w_j}^{1/2}$ and represented the matrix elements in numerical form (Table 4(a)).

**Table 4.** (**a**) Spinon flavor rotation matrix; (**b**) CKM matrix

| | $(u \vee d)$ | $(c \vee s)$ | $(t \vee b)$ |
|---|---|---|---|
| $(u \vee d)$ | 0.977 | 0.214 | 0.010 |
| $(c \vee s)$ | 0.214 | 0.976 | 0.046 |
| $(t \vee b)$ | 0.010 | 0.046 | 0.999 |

(**a**)

| | $d$ | $s$ | $b$ |
|---|---|---|---|
| $u$ | 0.974 | 0.222 | 0.004 |
| $c$ | 0.221 | 0.987 | 0.041 |
| $t$ | 0.008 | 0.039 | 1.013 |

(**b**)

Since the comparison of two matrices is evaluative, the numbers in both matrices are rounded to three significant digits, which is sufficient for this purpose.

## 14. Conclusion

We have described a possible scenario of vacuum self-organization, which results in the formation of a system of fields and particles resembling those of the standard model. The initial vacuum represents a primitive stochastic environment in the form of self-interacting vacuum dust. The space is regular 3D. No metric is assigned before self-organization. The medium is dissipative, far from thermodynamic equilibrium and driven by two mutually opposite forces. The form of emergent vacuum organization is a coherent time-crystal network. The model was formalized with the use of iterated maps, which possess Feigenbaum universality. Period-doubling bifurcations here were associated with time-related phase transitions, where the time-crystal network and its elements (vacuum vortices) undergo essential changes in their properties. We have described various phenomena that can emerge in this environment, including some that are usually associated with quantum and/or relativistic media. We can mention discrete spectra (ground states, charges, oscillation periods), link flavors, spatial phase coherency, virtual states, tunneling, entanglement, time-related uncertainty of states, time dilation, etc.

Two types of time intervals that emerge in dissipative periodical systems to which time crystal networks relate are the oscillation period and dissipation time constant. These intervals can be used as etalons for two different time scales. One scale is globally fixed, essentially quantum, and can be compared to the Plank time scale. The second scale varies from place to place, depends on the local degree of network excitation, and can be compared with the Einstein proper time in GR. Remarkably, they coexist peacefully in the described environment.

The network phase entrainment (synchronization) is responsible for the emergence of a phase-difference field. Phase transitions with period doubling diversify this field. The daughter fields differ in the number of degrees of freedom, different topologies of attractors and the strength of interactions. But all these different fields arise due to the same forces acting in the primary vacuum dust. In this regard, they can be considered as discrete states of the same (gravitation-antigravitation) field that are characterized by the index of their attractors $\mathcal{W}$ as a quantum number.

Elementary particles are represented by pairs of vortices with spinorial properties (spinons).

The diversity of spinons is sufficient to synthesize a family of particles whose taxonomy and organization resemble those of SMs. The spinon flavor set resembles that of quarks, and according to our estimates, the probabilities of their transmutations correspond with matrix elements of the CKM-matrix.

Due to the Feigenbaum universality and universal properties of time crystals, the approach and results can find applications in the study of other self-organizing systems.

**Conflicts of interest:** The author declares no conflicts of interest.

**Funding:** This research received no external funding.